%% file: Paper.tex
\newcommand*{\ATLASLATEXPATH}{latex/}
\documentclass[cernpreprint,texlive=2011,txfonts,UKenglish]{latex/atlasdoc}

\usepackage{\ATLASLATEXPATH atlaspackage}
\usepackage{\ATLASLATEXPATH atlasbiblatex}

\usepackage{\ATLASLATEXPATH atlascontribute}

\usepackage{\ATLASLATEXPATH atlasphysics}

\graphicspath{{logos/}{figures/}}

\usepackage{Paper-defs}

\newcommand{\pb}{\rm{pb}^{-1}}

\def\Zmm{{Z\to\mu\mu}}
\def\Jpsimm{{J/\psi\to\mu\mu}}

\input{Paper-metadata}

\hypersetup{pdftitle={ATLAS draft},pdfauthor={The ATLAS Collaboration}}

\begin{document}
\nolinenumbers

\maketitle

\section{Introduction}

Muons are key to some of the most important physics results published by the ATLAS experiment~\cite{atlas-det} at the LHC.
These results include the discovery of the Higgs boson~\cite{Aad:2012tfa} and  the measurement of
its properties~\cite{Higgs-mass, HiggsCPRun1, HiggsCouplingsRun1},
the precise measurement of  Standard Model processes~\cite{Aad:2015auj,Aad:2015zqe},
and  searches for physics beyond the Standard Model~\cite{Run1Zprime,Run1Wprime,Aad:2015ipg,Aad:2015sva}.

The performance of the ATLAS muon reconstruction during the LHC run at $\sqrt{s}$ = 7--8~\TeV\ has been documented in recent publications~\cite{perf2012,perf2010}.
During the 2013--2015 shutdown, the LHC was upgraded to increase the centre-of-mass energy from 8 to 13~\TeV\
and the ATLAS detector was equipped with additional muon chambers and
 a new innermost Pixel layer,  the Insertable B-Layer,  providing measurements closer to the interaction point.
Moreover, the muon reconstruction software was updated and improved.

After introducing the  ATLAS  muon reconstruction and identification algorithms,
this article  describes the performance of the  muon reconstruction
 in the first dataset collected  at $\sqrt{s}=13$~\TeV.
Measurements of the muon reconstruction and isolation efficiencies
and of the momentum scale and resolution are presented.
The comparison between data and Monte Carlo (MC) simulation and the determination of
the corrections to the simulation used in physics analyses are also discussed.
The results are based on the analysis of a large  sample of $\Jpsimm$ and $\Zmm$  decays reconstructed in
 3.2 \ifb\ of $pp$ collisions  recorded in 2015.

This article is structured as follows: Section~\ref{Sec:Detector} describes the  ATLAS subdetectors that are most relevant to this work;
Sections~\ref{Sec:MuonReco} and~\ref{Sec:MuonID} describe the muon reconstruction and identification in ATLAS, respectively;
Section~\ref{Sec:Datasets} describes the data samples used in the analysis;
the reconstruction and isolation efficiencies are described in Sections~\ref{Sec:Efficiency} and~\ref{Sec:Isolation}, respectively, while
the momentum scale and resolution  are described in \refS{Sec:Resolution}.
Finally, conclusions are given in \refS{sec:conclusions}.

\section{ATLAS detector}
\label{Sec:Detector}

A detailed description of the ATLAS detector can be found in Ref.~\cite{atlas-det}.
Information primarily from the inner detector (ID) and the muon spectrometer (MS), supplemented by information from  the calorimeters,
is used  to identify and precisely reconstruct muons produced in  $pp$ collisions.

The ID consists of three subdetectors: the silicon pixels (Pixel) and the semi\-con\-ductor tracker (SCT)
with a pseudorapidity\footnote{ATLAS uses a right-handed coordinate system with its origin at the nominal interaction point (IP) in the centre of the detector and the $z$-axis along the beam pipe. The $x$-axis points from the IP to the centre of the LHC ring, and the $y$-axis points upward. Cylindrical coordinates $(r,\phi)$ are used in the transverse plane, $\phi$ being the azimuthal angle around the beam pipe. The pseudorapidity and the transverse momentum are defined in terms of the polar angle $\theta$ as $\eta=-\ln\tan(\theta/2)$ and $\pt= p \sin \theta$, respectively. The $\eta$--$\phi$ distance between two particles is defined as  \mbox{$\Delta R=\sqrt{(\Delta\eta)^2 +(\Delta\phi)^2}$}.}
coverage up to $|\eta |$ = 2.5, and the transition radiation tracker (TRT) with a pseudorapidity coverage up to $|\eta |$ = 2.0.
The ID measures the muon track close to the interaction point, providing
accurate measurements of the track parameters inside an axial magnetic field of $2$~T.

The MS is the outermost ATLAS subde\-tector. It is designed to detect muons in the pseudorapidity  region up to $|\eta| = 2.7$,
and to provide momentum measurements with a relative resolution better than 3\% over a wide $\pt$ range and up to 10\% at $\pt \approx 1$~\TeV.
The MS consists of one barrel ($|\eta| < 1.05$) and two endcap sections ($1.05 < |\eta| < 2.7$). A system of three large superconducting air-core toroidal magnets,
each with eight coils, provides a magnetic field with a bending integral of about $2.5$~Tm in the barrel and up to $6$~Tm in the endcaps.
Resistive plate chambers (RPC, three doublet layers for $|\eta| < 1.05$) and  thin gap chambers (TGC,
one triplet layer followed by two doublets for
$1.0<|\eta| < 2.4$) provide triggering capability to the detector as well as  ($\eta$, $\phi$) position measurements with typical
spatial resolution of $5-10$~mm.
A precise  momentum measurement for muons with pseudorapidity up to $|\eta|=2.7$  is provided by three layers of monitored
drift tube chambers (MDT), with each chamber providing six to eight $\eta$ measurements along the muon trajectory.
For $|\eta|>2$,
the inner layer is instrumented with a quadruplet of cathode strip chambers (CSC) instead of MDTs.
The single-hit resolution in the bending plane for the MDT and the CSC is about $80$~$\mu$m and $60$~$\mu$m, respectively.
The muon chambers are aligned with a precision between $30$~$\mu$m  and $60$~$\mu$m.

During the shutdown preceding the LHC Run 2, the MS was completed to its initial design~\cite{ATLAS:1997ad}
by adding the last missing chambers in the transition region between the barrel and the endcaps ($1.0<|\eta|<1.4$).
Four RPC-equipped MDT chambers were also installed
inside two elevator shafts to improve the acceptance in that region compared to Run 1. Some of the new MDT chambers are made of tubes with a smaller radius compared to the ones used in the rest of the detector, allowing the detector to cope with higher rates.

The material between the interaction point (IP) and the MS ranges approximately from 100 to 190 radiation lengths, depending on $\eta$,
and consists mostly of  calorimeters.
The  lead/liquid-argon electromagnetic calorimeter covers $|\eta|<3.2$.
It is surrounded by hadronic calorimeters made of steel and scintillator tiles for $|\eta|< 1.7$,
 and copper or tungsten and liquid argon  for $|\eta|> 1.7$.

\section{Muon reconstruction }
\label{Sec:MuonReco}

Muon reconstruction is first performed independently in the ID and MS.
The information from individual subdetectors is then combined to form the muon tracks that are used in physics analyses.
In the ID, muons are reconstructed like any other charged particles as
described in Refs.~\cite{Cornelissen:1020106,ATLAS-CONF-2010-072}.
This section focuses on the description of the muon reconstruction in the MS (Section~\ref{Sec:MSmuon})
and on the combined muon reconstruction (Section~\ref{Sec:GlobalMuon}).

\subsection{Muon reconstruction in the MS}
\label{Sec:MSmuon}

Muon reconstruction in the MS starts with a search for hit patterns inside each
muon chamber to form segments.
In each MDT chamber and nearby trigger chamber, a Hough transform~\cite{Illingworth198887} is used to search for
hits aligned on a trajectory in the bending plane of the detector.
The MDT segments are reconstructed by performing a straight-line fit to the hits found in each layer.
The RPC or TGC hits  measure the coordinate orthogonal to the bending plane.
Segments in the CSC detectors are built using a separate combinatorial search in the $\eta$ and $\phi$ detector planes.
The search algorithm includes
a loose requirement on the compatibility of the track with the luminous region.

Muon track candidates are then built by fitting together hits from segments in different layers.
The algorithm used for this task
performs a segment-seeded combinatorial search that
starts by using as seeds the segments generated in the middle layers of the detector where more  trigger hits are available.
The search is then extended to use the segments from the outer  and  inner layers as seeds.
The segments are selected  using criteria based on hit multiplicity and fit quality and
are matched using their relative positions and angles.
At least two matching segments are required to build a track, except in the barrel--endcap transition region
where a single high-quality segment with $\eta$ and $\phi$ information can be used to build a track.

The same segment can initially be used to build several track candidates.
Later, an overlap removal algorithm selects the best assignment to a single track,
or allows for the segment to be shared
between two tracks.
To ensure  high efficiency for close-by muons,
all  tracks with segments in three different layers of the spectrometer are kept when they are identical in two out of three layers
but share no hits in the outermost layer.

The hits associated with each  track candidate are fitted using a global $\chi^2$ fit.
A track candidate is accepted if the $\chi^2$ of the fit satisfies the selection criteria.
Hits providing large contributions to the $\chi^2$ are removed and the track fit is repeated.
A hit recovery procedure is also performed looking for additional hits consistent with the candidate trajectory.
The track candidate is refit if additional hits are found.

\subsection{Combined reconstruction }
\label{Sec:GlobalMuon}

The combined ID--MS muon reconstruction is performed according to various
algorithms based on the information provided by the ID,  MS, and  calorimeters.
Four muon {\em types} are defined depending on which subdetectors are used in reconstruction:
\begin{itemize}
\item Combined (CB) muon: track reconstruction is performed independently in the ID and MS, and a combined track
  is formed with a global refit that uses the hits from both the ID and MS subdetectors.
  During the global fit procedure, MS hits may be added to or removed from the track to improve the fit quality.
  Most muons are reconstructed  following an outside-in pattern recognition, in which the muons are first reconstructed in the MS and then extrapolated inward and matched to an ID track.
  An inside-out combined reconstruction, in which ID tracks are extrapolated outward and matched to MS tracks, is  used as a complementary approach.

\item Segment-tagged (ST) muons: a track in the ID is classified as a muon if, once extrapolated to the MS, it is associated
  with at least one local track segment in the MDT or CSC chambers.
  ST muons are used
  when muons cross only one layer of MS chambers, either because of their low $\pt$ or because they fall in regions with  reduced MS acceptance.

\item Calorimeter-tagged (CT) muons: a track in the ID is identified as a muon if it can be matched to an energy deposit
  in the calorimeter compatible with a minimum-ionizing particle. This type has the lowest purity of all the muon types but it
  recovers acceptance in the region
  where the ATLAS muon spectrometer is only partially instrumented to allow for cabling and services to the calorimeters and inner detector.
  The identification criteria for CT  muons are optimised for
  that region ($|\eta|<0.1$) and a momentum range of $15< \pt < 100$~\GeV.

\item Extrapolated (ME) muons: the muon trajectory is reconstructed based only on the MS track
  and
      a loose requirement on compatibility with originating from  the IP.
  The parameters of the muon track are defined at the interaction point,
  taking into account
  the estimated energy loss of the muon in the calorimeters.
  In general, the muon is required to traverse at least two layers of MS chambers to provide a track measurement, but three layers are required
  in the forward region.
  ME muons are mainly used to extend the acceptance for muon reconstruction into the region $2.5<|\eta|<2.7$, which  is not covered by the ID.

\end{itemize}

Overlaps between different muon types are resolved before producing the collection of muons used in physics analyses.
When two muon types share the same ID track, preference is given to CB muons, then to ST, and finally to CT muons.
The overlap with ME muons in the muon system is resolved by analyzing the track hit content and selecting the track with better fit quality and
larger number of hits.

The muon reconstruction used in this work evolved from the algorithms
defined as  {\em Chain 3} in Ref.~\cite{perf2012}.
These algorithms were improved in several ways.
The use of a Hough transform
to identify the hit patterns for seeding the segment-finding algorithm
makes the reconstruction faster and more robust against misidentification of hadrons,
thus providing better background rejection early in the pattern recognition process.
The calculation of the energy loss in the calorimeter was also improved.
An  analytic parameterization of the average energy loss is derived
from  a  detailed description of the detector geometry.
The final estimate of the energy loss is obtained by combining the
analytic parameterization  with the energy measured in the calorimeter.
This method yields a precision on the mean energy loss of about  30 MeV
for 50\,\GeV\ muons.

\section{Data and Monte Carlo samples}
\label{Sec:Datasets}

The  efficiency measurements presented in this article
are obtained from the analysis of 3.2~\ifb\ of $pp$ collision
data recorded at $\sqrt{s}=13$~TeV at the LHC in 2015 during the
data-taking period with  25 ns spacing between bunch crossings.
About  $1.5\,{\rm M}$ $\Zmm$  and $3.5\,{\rm M}$  $\Jpsimm$ events are reconstructed and used for the analysis.
For the study of the momentum calibration, 2.7~\ifb\ of data were used, rejecting
the runs in which the longitudinal position of the beam spot was displaced by about 3~cm with respect to the centre of the detector.

Events are accepted only if the ID, the MS, and
the calorimeters  were operational and
the solenoid and toroid magnet systems were both active.
The online event selection was performed by a two-level trigger system derived from the one described in Ref.~\cite{Aad:2012xs}.
The $Z\to\mu\mu$ candidates  are triggered by the presence of at least one muon candidate with a
transverse momentum,
\pt , of at least 20~GeV. For the reconstruction efficiency and momentum calibration studies, the muon firing the trigger is required
to be isolated (see Section~\ref{Sec:Isolation}).
The $\Jpsimm$  candidates used for the momentum calibration are triggered by a dedicated dimuon trigger
that requires two opposite-charge muons, each with $\pt>4$~GeV,
com\-pa\-tible with the same vertex,
and with a dimuon invariant mass in the range 2.5--4.5 GeV.
The $\Jpsimm$ sample used for the efficiency measurement
is selected using a combination  of single-muon triggers
and  triggers requiring one muon with transverse momentum of at least 4~GeV and an ID track such that the invariant mass of
the muon+track pair, under a muon mass hypothesis, is compatible with the mass of the $J/\psi$.

Monte Carlo samples for the process $pp \rightarrow (Z/\gamma^*) X \rightarrow \mu\mu X$ are generated
using the POWHEG BOX~\cite{Alioli:2010xd} interfaced to PYTHIA8 \cite{pythia8} and
the CT10~\cite{Lai:2010vv} parton distribution functions.
The PHOTOS~\cite{Golonka:2005pn} package
is used to simulate final-state photon radiation in $Z$ boson decays.
Samples of prompt $\Jpsimm$ decays are generated using PYTHIA8 complemented with PHOTOS to simulate the effects of final-state radiation.
A requirement on the minimum transverse momentum of each  muon  ($\pt>4$\,\GeV ) is applied at the generator level.
The samples used for the simulation of the backgrounds to $\Zmm$
include: $Z\to\tau\tau$, $W\to\mu\nu$, and $W\to\tau\nu$, generated with POWHEG BOX; $WW$, $ZZ$, and $WZ$ generated with SHERPA~\cite{sherpa};
$t\bar{t}$ samples generated with POWHEG BOX + PYTHIA8;
and $b\bar{b}$ and  $c\bar{c}$ samples generated with PYTHIA8.

All the generated samples are passed through the simulation of the ATLAS detector based on GEANT4~\cite{simuAtlas,GEANT4}
and are reconstructed with the same programs used for the data.
The ID and the MS are simulated with an ideal geometry assuming no misalignment.

The effect of multiple $pp$ interactions per bunch crossing (``pile-up'')
is modelled by overlaying simulated minimum-bias events onto the
original hard-scattering event. Monte Carlo events are then reweighted so that
the distribution of the average number of interactions per event agrees with the data.

\section{Muon identification}
\label{Sec:MuonID}

Muon identification is performed by applying  quality requirements
that suppress background,  mainly from pion and kaon decays,
while selecting prompt muons with high efficiency  and/or guaranteeing a robust momentum measurement.

Muon candidates originating from in-flight decays of charged hadrons in the ID are often characterized by the presence of a distinctive ``kink'' topology in the reconstructed track.
As a  consequence, it is expected that the fit quality of the resulting combined track will be  poor and that
the momentum measured in the ID and MS may  not be compatible.
 Several variables offering good discrimination between prompt muons and background muon candidates are studied in simulated $t\bar{t}$ events.
Muons from $W$ decays are categorized as {\em signal} muons
while  muon candidates from light-hadron decays are categorized as {\em background}.
For CB tracks, the   variables used in  muon identification are:
\begin{itemize}

\item  {\em q/p significance}, defined as the absolute value of the difference between the ratio of the charge and momentum of the muons measured
in the ID and  MS divided by the sum in quadrature of the corresponding uncertainties;

\item $\rho^{\prime}$, defined as the absolute value of the difference between the transverse momentum measurements
in the ID and MS divided by the \pt\ of the combined track;

\item normalised $\chi^2$  of the combined track fit.

\end{itemize}

To guarantee a robust momentum measurement, specific requirements on the number of hits in the ID and MS are used.
For the ID, the quality cuts require
at least one Pixel hit,
at least five SCT hits,
fewer than three Pixel or SCT holes,
and that at least 10\% of the TRT hits originally assigned to the track are included in the final fit;
the last requirement is only employed for $|\eta|$ between 0.1 and 1.9, in the region of full TRT acceptance.
A hole is defined as  an active sensor traversed by the track but containing no hits.
A missing hit is considered a hole only when it falls  between  hits successfully
assigned to a given track. If some inefficiency is expected for a given sensor, the requirements  on the number of Pixel and SCT hits are reduced accordingly.

Four muon identification selections
(\emph{Medium}, \emph{Loose},  \emph{Tight}, and \emph{High-p$_{\rm T}$})
are  provided to address the specific needs of different physics analyses.
\emph{Loose}, \emph{Medium}, and \emph{Tight} are inclusive categories in that muons identified with tighter requirements are also
included in the looser categories.

{\bf \emph{Medium} muons }
The \emph{Medium} identification criteria   provide the default selection for muons in ATLAS.
This selection minimises the systematic uncertainties associated with muon reconstruction and calibration.
Only CB and ME tracks are used.
The former are required to have $\geq 3$ hits in at least two MDT layers, except for tracks in the $|\eta|<0.1$ region,
where tracks with at least one MDT layer but no more than one MDT hole layer are allowed.
The latter are required to have at least three MDT/CSC layers, and are employed only in the $2.5<|\eta|<2.7$ region to extend the acceptance outside the ID geometrical coverage.
A loose selection on the compatibility between ID and MS momentum measurements is applied to suppress the contamination
due to hadrons misidentified as muons. Specifically, the {\em q/p significance}  is required to be less than seven.
In the pseudorapidity region $|\eta|<2.5$, about
0.5\% of the muons classified as \emph{Medium}  originate from the
inside-out combined reconstruction strategy.

{\bf \emph{Loose} muons }
 The \emph{Loose} identification criteria are designed to maximise the reconstruction efficiency while providing good-quality muon tracks.
They are specifically optimised for reconstructing Higgs  boson candidates in the four-lepton final state~\cite{HiggsCouplingsRun1}. All  muon types are used.
All CB and ME muons satisfying the \emph{Medium} requirements are included in the \emph{Loose} selection.
CT and ST muons are restricted to the $|\eta|<0.1$ region.
In the region $|\eta|<2.5$, about  97.5\% of the \emph{Loose} muons are combined muons,
approximately 1.5\% are CT and the remaining 1\% are reconstructed as ST muons.

{\bf \emph{Tight} muons }
\emph{Tight} muons are selected to maximise the purity of muons at the cost of some efficiency.
Only CB muons with hits in at least two stations of the MS and satisfying the \emph{Medium} selection criteria are considered.
The normalised $\chi^2$  of the combined track fit is required to be $<8$  to remove pathological tracks.
A two-dimensional cut in the  $\rho^{\prime}$ and  {\em q/p significance} variables is performed as a function of the muon \pt\ to ensure
 stronger background rejection for momenta below 20~\GeV\ where the misidentification probability is higher.

{\bf \emph{High-p$_{\rm T}$}  muons }
The \emph{High-p$_{\rm T}$} selection aims to maximise the momentum resolution for tracks with transverse momentum
above 100~\GeV.
The selection is optimised for searches for high-mass $Z'$ and $W'$ resonances~\cite{Run1Zprime,Run1Wprime}.  
CB muons passing the \textit{Medium} selection and having at least  three hits in three MS stations are selected.
Specific regions of the MS where the alignment is suboptimal are vetoed as a precaution.
Requiring three MS stations, while  reducing the reconstruction efficiency by about 20\%,
improves the $\pt$ resolution of muons above 1.5~\TeV\ by approximately 30\%.

The reconstruction efficiencies for {\em signal}
and {\em background}
 obtained from  $t\overline{t}$ simulation are reported in Table~\ref{tab:rates_25ns}.
The results are shown for the four identification selection criteria separating  low ($4<\pt<20$~\GeV) and high  ($20<\pt<100$~\GeV) transverse momentum muon candidates.
No isolation requirement is applied in the selection shown in the table. When isolation requirements are applied,
the misidentification rates are reduced by more than an order of magnitude.
It should be noted that the higher misidentification rate observed for {\em Loose}  with respect to  {\em Medium} muons is mainly due to CT muons in the region $|\eta |<0.1$.

The misidentification probability estimated with the MC simulation is validated in data
by measuring the probability that pions are reconstructed as muons. An unbiased sample of pions
from $K_{\rm S}^0\to\pi^+\pi^-$  decays is collected with calorimeter-based (photon, electron, jet) triggers.
Good agreement between data and simulation is observed independent of the \pt , $\eta$, and impact parameter of the track.

\begin{table*}[!h]
 \renewcommand{\arraystretch}{1.2}
\begin{center}
\begin{tabular}{|c||c|c||c|c|}
\hline
\hline
               &   \multicolumn{2}{c||}{ $4<\pt<20$~\GeV}       &      \multicolumn{2}{c|}{ $20<\pt<100$~\GeV}    \\

\hline
 Selection  &  $\epsilon_{\mu}^{\mathrm{MC}}$ [\%] &  $\epsilon_{\mathrm{Hadrons}}^{\mathrm{MC}}$ [\%] &  $\epsilon_{\mu}^{\mathrm{MC}}$ [\%] &  $\epsilon_{\mathrm{Hadrons}}^{\mathrm{MC}}$ [\%]  \\

\hline
Loose             & 96.7   & 0.53 & 98.1 & 0.76 \\
Medium            & 95.5   & 0.38 & 96.1 & 0.17 \\
Tight             & 89.9   & 0.19 & 91.8 & 0.11 \\
High-p$_{\rm T}$   & 78.1   & 0.26 & 80.4 & 0.13 \\
\hline
\hline
\end{tabular}
\end{center}
\caption{\label{tab:rates_25ns} Efficiency for prompt muons from $W$ decays and hadrons decaying in flight and misidentified as prompt muons
computed using a $t\bar{t}$ MC sample. The results are shown for the four identification selection criteria
separating  low ($4<\pt<20$~\GeV) and high  ($20<\pt<100$~\GeV) momentum muons
for candidates with $|\eta |<2.5$. The statistical uncertainties are negligible. }
\end{table*}

\section{Reconstruction efficiency}
\label{Sec:Efficiency}

As the muon reconstruction in the ID and MS detectors is performed independently, a precise
determination of the muon reconstruction efficiency
in the region $|\eta|<2.5$ is obtained with the tag-and-probe method,
as described in the Section~\ref{sec::muon_reco_central}.
A different methodology, described in Section~\ref{sec::muon_reco_forward},
is used in the region  $2.5<|\eta|<2.7$ where muons are reconstructed  using only the MS detector.

\subsection{Efficiency measurement in the region $|\eta|<2.5$}
\label{sec::muon_reco_central}

The tag-and-probe method
is employed to measure the efficiency of the muon identification selections within the acceptance of the ID ($|\eta|<2.5$).
The method is based on the selection of an almost pure muon sample from $\Jpsimm$ or $\Zmm$ events,
requiring one leg of the decay (tag) to be identified as a \emph{Medium} muon that fires the trigger and
the second leg (probe) to be reconstructed
by a system independent of the one being studied.
A selection based on the event topology is used to reduce the background contamination.

Three kinds of probes are used to measure muon efficiencies.
ID tracks and CT muons both allow a measurement of the efficiency in the MS,
while MS tracks are used to determine the complementary efficiency of the muon reconstruction in the ID.
Compared to ID tracks, CT muons offer a more powerful rejection of backgrounds, especially at low transverse momenta,
and are therefore the preferred probe type for this part of the measurement.
ID tracks are used as a cross-check and for measurements not directly accessible to CT muons.
A direct measurement of the CT muon reconstruction efficiency is possible using MS tracks.

The efficiency measurement for \emph{Medium}, \emph{Tight}, and \emph{High-p$_{\rm T}$}   muons consists of two stages.
First, the efficiency $\epsilon\left(\textrm{X} | \textrm{CT}\right)$ (X = \emph{Medium} / \emph{Tight} / \emph{High-p$_{\rm T}$})
of reconstructing these muons assuming a reconstructed ID track is measured using a CT muon as probe.
Then, this result is corrected by the efficiency $\epsilon\left( \textrm{ID} | \textrm{MS} \right)$ of the ID
track reconstruction, measured using MS probes:
\begin{equation}
    \epsilon\left(\textrm{X}\right) = \epsilon\left(\textrm{X} | \textrm{ID} \right) \cdot \epsilon \left(\textrm{ID}\right)
=
\epsilon\left(\textrm{X} | \text{CT} \right) \cdot \epsilon \left(\textrm{ID} | \textrm{MS}\right) \phantom{,,,} (\textrm{X} = \textrm{\emph{Medium}~/~\emph{Tight}~/~\emph{High-p$_{\rm T}$}}).
\label{eq:effTPformula}
\end{equation}
A similar approach is used when using ID probe tracks for cross-checks.

This approach is valid if two assumptions are satisfied:
\begin{itemize}
\item the ID track reconstruction efficiency is independent from the muon
  spectrometer track reconstruction ($\epsilon\left( \textrm{ID}\right) =  \epsilon\left( \textrm{ID} | \textrm{MS} \right)$).
\item the use of a CT muon
  as a probe instead of an ID track does not affect the probability for \emph{Medium}, \emph{Tight}, or \emph{High-p$_{\rm T}$}
  reconstruction  ($\epsilon\left( \textrm{X} | \textrm{ID} \right) =  \epsilon\left(  \textrm{X} | \textrm{CT}  \right)$).
\end{itemize}
Both assumptions have been tested using generator-level information from simulation and small differences are taken into account in the
systematic uncertainties.

The muons selected by the \emph{Loose} identification requirements are decomposed into two samples:
CT muons within $|\eta| < 0.1$ and all other muons.
The CT muon efficiency is measured using MS probe tracks,
while the efficiency of other muons is evaluated using CT probe muons in a fashion similar
to the \emph{Medium}, \emph{Tight}, and \emph{High-p$_{\rm T}$} categories.

The level of agreement of the measured efficiency, $\epsilon^{{\textrm{Data}}}\left(\textrm{X}\right)$, with the efficiency measured
with the same method in simulation, $\epsilon^{{\textrm{MC}}}\left(\textrm{X}\right)$, is expressed as the ratio of these two numbers,
called the ``efficiency scale factor'' (\textrm{SF}):
\begin{equation}
  \textrm{SF}=\frac{\epsilon^{{\textrm{Data}}}\left(\textrm{X}\right)}{\epsilon^{{\textrm{MC}}}\left(\textrm{X}\right)} \textrm{.}
\end{equation}
This quantity describes the deviation of the simulation from the real detector behaviour, and is of particular
interest to physics analyses, where it is used to correct the simulation.

\subsubsection{The tag-and-probe method with $\Zmm$ events}
\label{sec::T&PZ}

Events are selected by requiring muon pairs with an invariant mass within $ 10\GeV$ of the $Z$ boson  mass.
The tag muon is required to satisfy the \emph{Loose} isolation (see Section~\ref{sec::muon_isol_perf})
and \emph{Medium} muon identification selections
and to have a transverse momentum of at least $24\GeV$.
Requirements on the significance of the transverse impact parameter $d_0$
($|d_0|/\sigma(d_0) < 3.0$)
and on the longitudinal impact parameter $|z_0|$  ($|z_0|$ < 10~mm) of the tag muon are imposed.
Finally, the tag muon is required to have triggered the readout of the event.

The probe muon is required to have a transverse momentum of at least $10\GeV$
and to satisfy the \emph{Loose} isolation criteria.
While this is sufficient to ensure  high purity in the case of MS probe tracks,
further requirements are applied to both the ID track and CT muon probes.
In the case of ID tracks, an isolation requirement is applied which is considerably
stricter than the \emph{Loose} selection in order to suppress backgrounds as much as possible.
In addition, the invariant mass window is tightened to $5\GeV$ around the $Z$ boson mass, rather than
the $10\GeV$ used in the other cases.
For CT muon probes, additional requirements on the compatibility of the associated
calorimeter energy deposit with a muon signature are applied to further enhance the purity.
The ID probe tracks and calorimeter-tagged probe muons must also have  transverse
and longitudinal impact parameters consistent with being produced in a primary $pp$ interaction, as required
for tag muons.
A probe is considered successfully reconstructed if a reconstructed muon is found within
a cone in the $\eta$--$\phi$ plane of size $\Delta R = 0.05$ around the probe track.

A small fraction (about 0.1\%) of the selected tag--probe pairs originates from sources other than $Z\to\mu\mu$ events.
For a precise efficiency measurement, these backgrounds must be estimated and subtracted.
Contributions from $Z\to\tau\tau$ and $t\bar{t}$ decays are estimated using simulation.
Additionally,  multijet events and $W\to\mu\nu$ decays in association with jet activity ($W$+jets) can yield tag--probe
pairs through secondary muons from heavy- or light-hadron decays. As these backgrounds are approximately charge-symmetric,
they are estimated from the data using same-charge (SC) tag--probe pairs.
This leads to the following estimate of the opposite-charge (OC) background,    $N^{\textrm{Bkg}}$, for each region of the kinematic phase-space:
\begin{equation}
  N^{\textrm{Bkg}} = N_{\text{OC}}^{Z,t\bar{t}\text{ MC}} + T \cdot \left( N_{\text{SC}}^{\text{Data}} - N_{\text{SC}}^{Z,t\bar{t}\text{ MC}} \right)
\end{equation}
where $N_{\text{OC}}^{Z,t\bar{t}\text{ MC}}$ is the contribution from  $Z\to\tau\tau$ and $t\bar{t}$ decays, $N_{\text{SC}}^{\text{Data}}$ is the number
of SC pairs measured in data and $N_{\text{SC}}^{Z,t\bar{t}\text{ MC}}$ is the estimated contribution of the  $Z\to\mu\mu$, $Z\to\tau\tau$, and $t\bar{t}$
processes to the SC sample.
$T$ is a global transfer factor that takes into account the charge asymmetry of the multijet and $W$+jets processes, estimated
in data using a control sample of events obtained by inverting the probe isolation requirement.
For MS (ID) tracks, a value of $T = 1.7$ $(1.1)$ is obtained, while for calorimeter-tagged muon probes the transfer factor is $T = 1.2$.
The systematic uncertainties in the transfer factor vary between 40\% and 100\% and are included in the systematic error in the reconstruction efficiency described
in Section~\ref{sec::T&Psys}.

The efficiency measured in the data is corrected for the background contributions described above
by subtracting the predicted probe yields attributed to these sources from the number of observed probes,
\begin{equation}
    \epsilon = \frac{N_{\textrm{R}}^{\textrm{Data}} - N_{\textrm{R}}^{\textrm{Bkg}}}{N_{\textrm{P}}^{\textrm{Data}} - N_{\textrm{P}}^{\textrm{Bkg}}},
\end{equation}
where $N_{\textrm{P}}$  denotes the total number of probes and $N_{\textrm{R}}$ the number of successfully reconstructed probes.
The resulting efficiency can then be compared directly to the result of the simulation.

\subsubsection{The tag-and-probe method with $\Jpsimm$ events}
\label{sec::T&PJPsi}

The reconstruction efficiencies of the \emph{Loose}, \emph{Medium}, and \emph{Tight} muon selections
at low $\pt$ are measured from a sample of $\Jpsimm$ events selected using a combination of single-muon triggers
and the dedicated ``muon + track'' trigger described in Section~\ref{Sec:Datasets}.

Tag--probe pairs are selected within the invariant mass window of $2.7$--$3.5\GeV$
and requiring a transverse momentum of at least $5\GeV$ for each muon.
The tag muon is required to satisfy the \emph{Medium} muon identification selection and
to have triggered the readout of the event.
In order to avoid low-momentum curved tracks sharing the same trigger region,
tag and probe muons are required to be $\Delta R >$ 0.2  apart when extrapolated
to the MS trigger surfaces.
Finally, they are selected with $\Delta z_0 \equiv |z_0^{\mathrm{tag}} - z_0^{\mathrm{probe}}| < 5$~mm, to
suppress background.
A probe is considered successfully reconstructed if a selected muon is found within
a $\Delta R$ = 0.05 cone around the probe track.

The background contamination and the muon reconstruction efficiency are measured with a simultaneous
maximum-likelihood fit of two statistically independent distributions of the invariant mass:
events in which the probe is or is not successfully matched to the selected muon.
The fits are performed in six $\pt$ and nine $\eta$ bins of the probe tracks.
The signal is modelled with a Crystal Ball function~\cite{crystalball2} with a single set of parameters for
the two independent samples.
Separate first-order polynomial fits are used to describe the background shape for
matched and unmatched probes.

\subsubsection{Systematic uncertainties}
\label{sec::T&Psys}
The main contributions to the systematic uncertainty in the measurement of the efficiency SFs
with $\Zmm$ and $\Jpsimm$ events are shown in Figs.~\ref{Fig:SF_syst_eta} and \ref{Fig:SF_syst_pt},
as a function of $\eta$ and $\pt$, respectively.

\begin{figure}[!hbp]
  \begin{center}
    \includegraphics[width=0.49\linewidth]{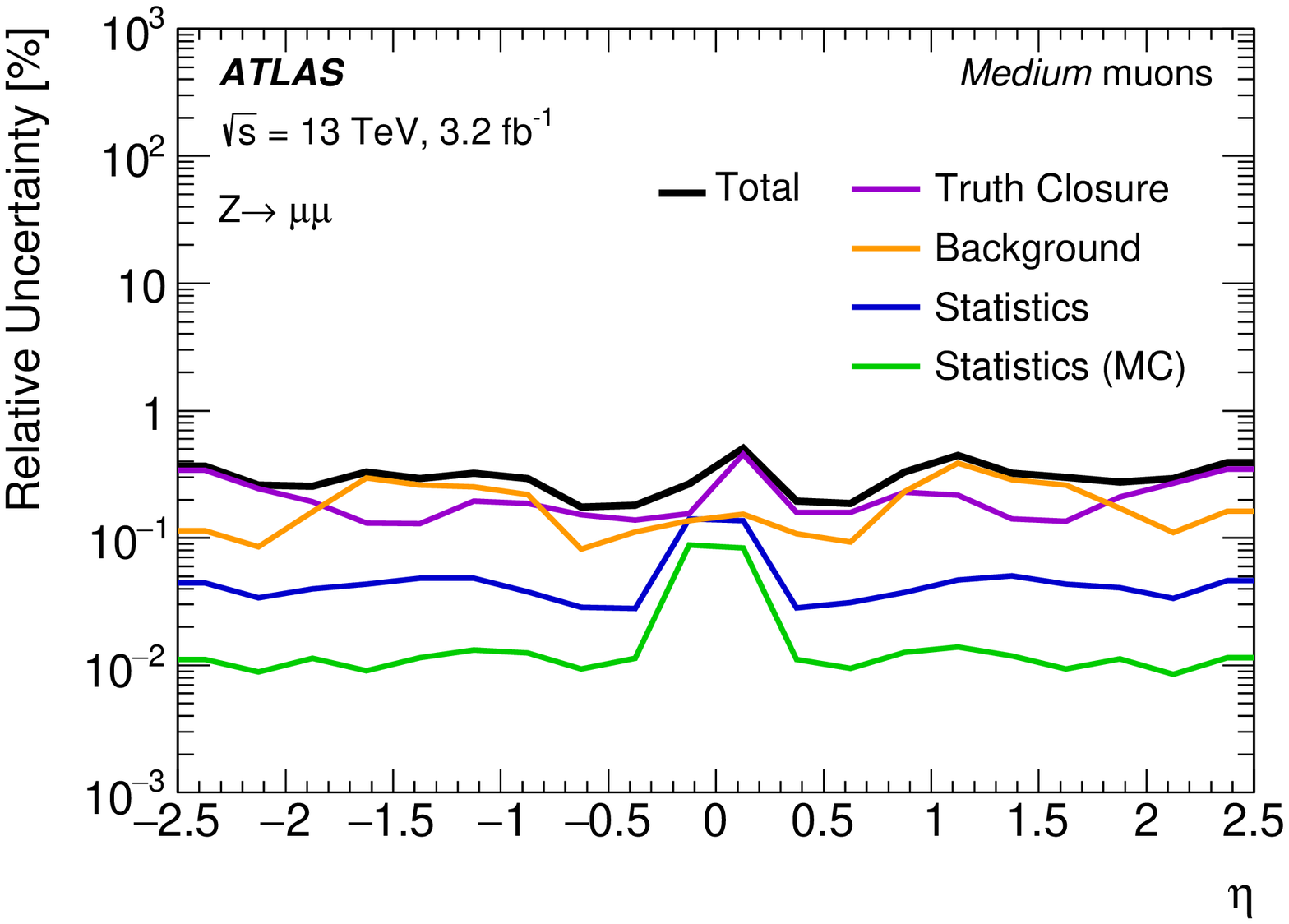}
    \includegraphics[width=0.49\linewidth]{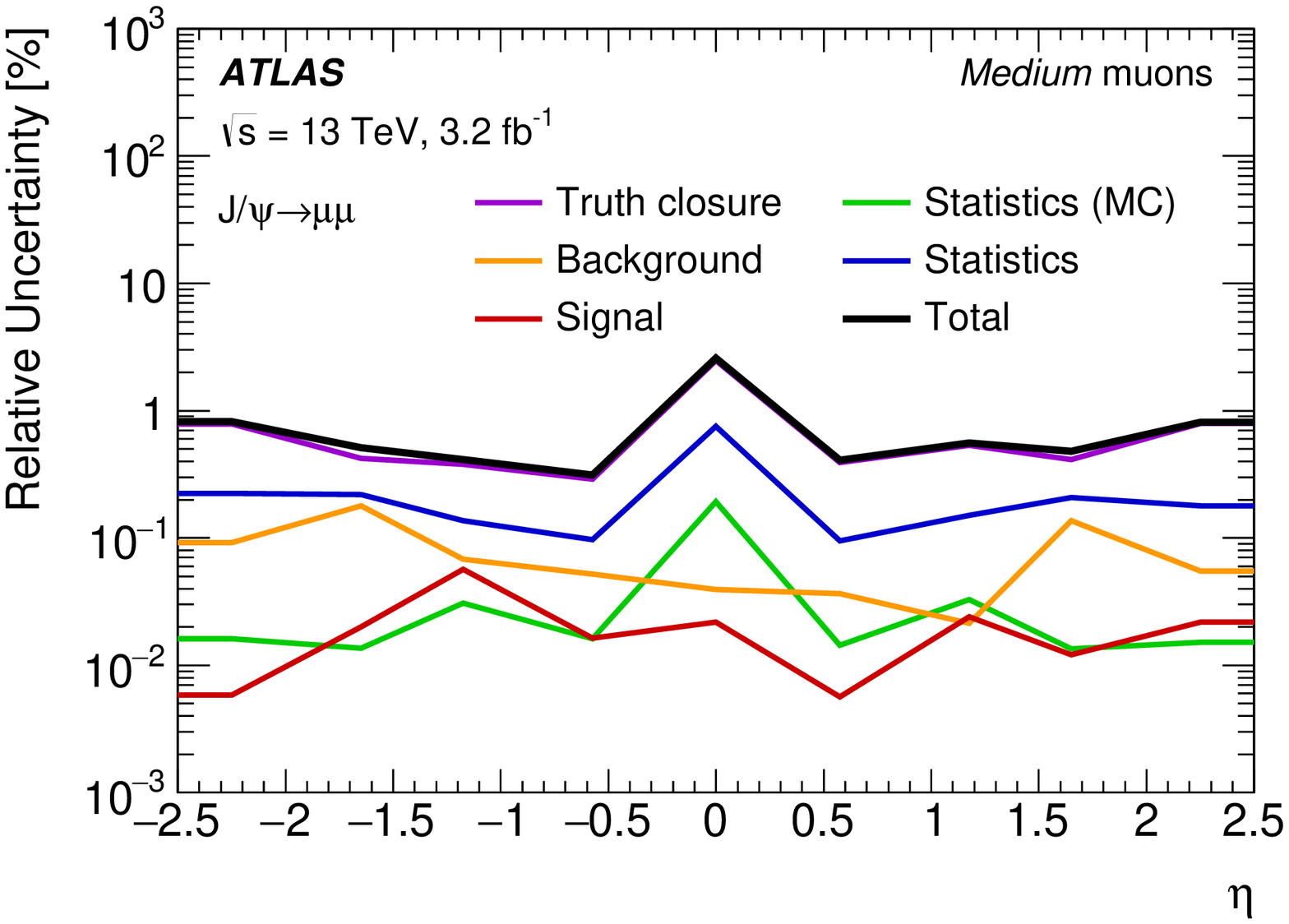}
    \caption{Total uncertainty in the efficiency scale factor for \emph{Medium} muons as a function of $\eta$ as
      obtained from $\Zmm$ data (left) for muons with $\pt>10$~\GeV, and from $\Jpsimm$ data (right) for muons with
      $5<\pt<20$~\GeV.
      The combined  uncertainty is the sum in quadrature of the individual contributions.}
    \label{Fig:SF_syst_eta}
\end{center}
\end{figure}

The uncertainty in the background estimate is evaluated in the $\Zmm$ analysis
by taking the maximum variation of the transfer factor $T$ when estimated with a
simulation-based approach as described in Ref.~\cite{perf2012} and when assuming the background
to be charge-symmetric. This results in an uncertainty of the efficiency measurement below $0.1\%$
over a large momentum range, but reaching $\sim 1\%$ for low muon momenta where the contribution
of the background is most significant.
In the $\Jpsimm$ analysis, the background uncertainty is estimated by changing the function used
in the fit to model the background, replacing the first-order polynomial with an exponential function.
An uncertainty due to the signal modelling in the fit, labelled as ``Signal'' in Figs.~\ref{Fig:SF_syst_eta} and~\ref{Fig:SF_syst_pt}, is also estimated using a convolution of exponential and
Gaussian functions as an alternative model. Each uncertainty is about $0.1\%$.

The cone size used for matching selected muons to probe tracks is optimised in terms of efficiency and purity of the matching.
The systematic uncertainty deriving from this choice is evaluated by varying the cone size by $\pm 50\%$.
This yields an uncertainty below $0.1\%$ in both analyses.

Possible biases in the tag-and-probe method,
such as biases due to different kinematic distributions between reconstructed
probes and generated
 muons or  correlations between ID and MS efficiencies, are estimated in simulation by
comparing the efficiency measured with the tag-and-probe method with the ``true'' efficiency
given by  the fraction of generator-level muons that are successfully reconstructed.
This uncertainty is labelled as ``Truth Closure'' in Figs.~\ref{Fig:SF_syst_eta} and \ref{Fig:SF_syst_pt}.
In the $\Zmm$ analysis, agreement better than $0.1\%$ is observed in the high momentum range.
This uncertainty grows at low $\pt$,  and differences up to $0.7\%$ are found in the $\Jpsimm$ analysis.
A larger effect of up to 1--2$\%$ is measured in both analyses in the region $|\eta|<0.1$.
In the extraction of the efficiency scale factors, the difference between the measured and the ``true''
efficiency cancels to first order. To take into account possible imperfections of the simulation, half of
the observed difference is used as an additional systematic uncertainty in the SF.

No significant dependence of the measured SFs with $\pt$ is observed in the momentum range considered
in the $\Zmm$ analysis.
An upper limit on the SF variation for large muon momenta is extracted from simulation,
leading to an additional uncertainty of 2--3\% per TeV for muons with $\pt>200$~\GeV.
The efficiency scale factor is observed to be independent of the amount of pile-up.

\begin{figure}[htbp]
  \begin{center}
    \includegraphics[width=0.49\linewidth]{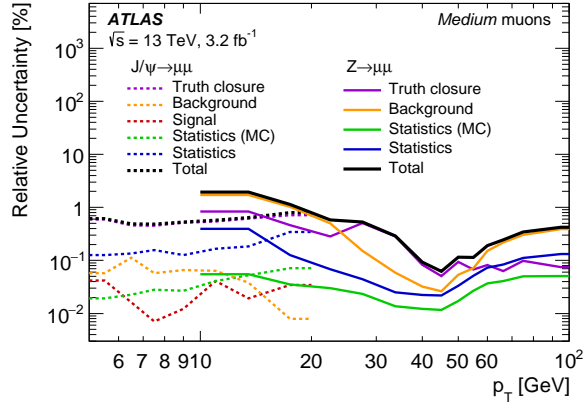}
    \caption{Total uncertainty in the efficiency scale factor for \emph{Medium} muons as a function of $\pt$ as
      obtained from $\Zmm$ (solid lines) and $\Jpsimm$ (dashed lines) decays.
      The combined  uncertainty is the sum in quadrature of the individual contributions.}
    \label{Fig:SF_syst_pt}
\end{center}
\end{figure}

\subsubsection{Results}
\label{sec::T&Pres}

Figure~\ref{Fig:AllTypeEff_Z} shows the muon reconstruction efficiency as a function of $\eta$ as measured from $\Zmm$ events
for the different muon selections. The efficiency as measured in data and the corresponding scale factors for the \emph{Medium}
selection are also shown in Fig.~\ref{Fig:Medium2DEff_Z} as a function of $\eta$ and $\phi$.
The efficiency at low $\pt$ is reported in Fig.~\ref{Fig:AllTypeEff_JPsi} as measured
from $\Jpsimm$ events as a function of $\pt$ in different $\eta$ regions.

The efficiencies of the \emph{Loose} and \emph{Medium} selections are very similar throughout  the detector with the exception of the
region  $|\eta|<0.1$, where the  \emph{Loose} selection fills the MS acceptance gap using the calorimeter and segment-tagged muons contributions.
The efficiency of these selections is observed to be in excess of $98\%$,
and between $90\%$ and $98\%$ for the \emph{Tight} selection,
with all efficiencies in very good agreement with those predicted by the simulation.
An inefficiency due to a poorly aligned MDT chamber is clearly localised at $(\eta , \phi) \sim (-1.3, 1.6)$, and is the most significant
feature of the comparison between collision data and simulation for these three categories.
In addition, a $2\%$-level local inefficiency is visible in the region
 $(\eta , \phi) \sim (1.9, 2.5)$,  traced to temporary failures in the SCT readout system.
Further local inefficiencies in the barrel region around $\phi \sim -1.1$ are also linked to temporary faults during data taking.
The efficiency of the  \emph{High-p$_{\rm T}$} selection is significantly lower, as a consequence of the strict requirements on momentum resolution.
Local disagreements between prediction and observation are more severe than in the case of the other muon selections.  Apart from
the poorly aligned MDT chamber, they are  most prominent in the CSC region.

\begin{figure}[bh]
\begin{center}
    \includegraphics[width=0.49\linewidth]{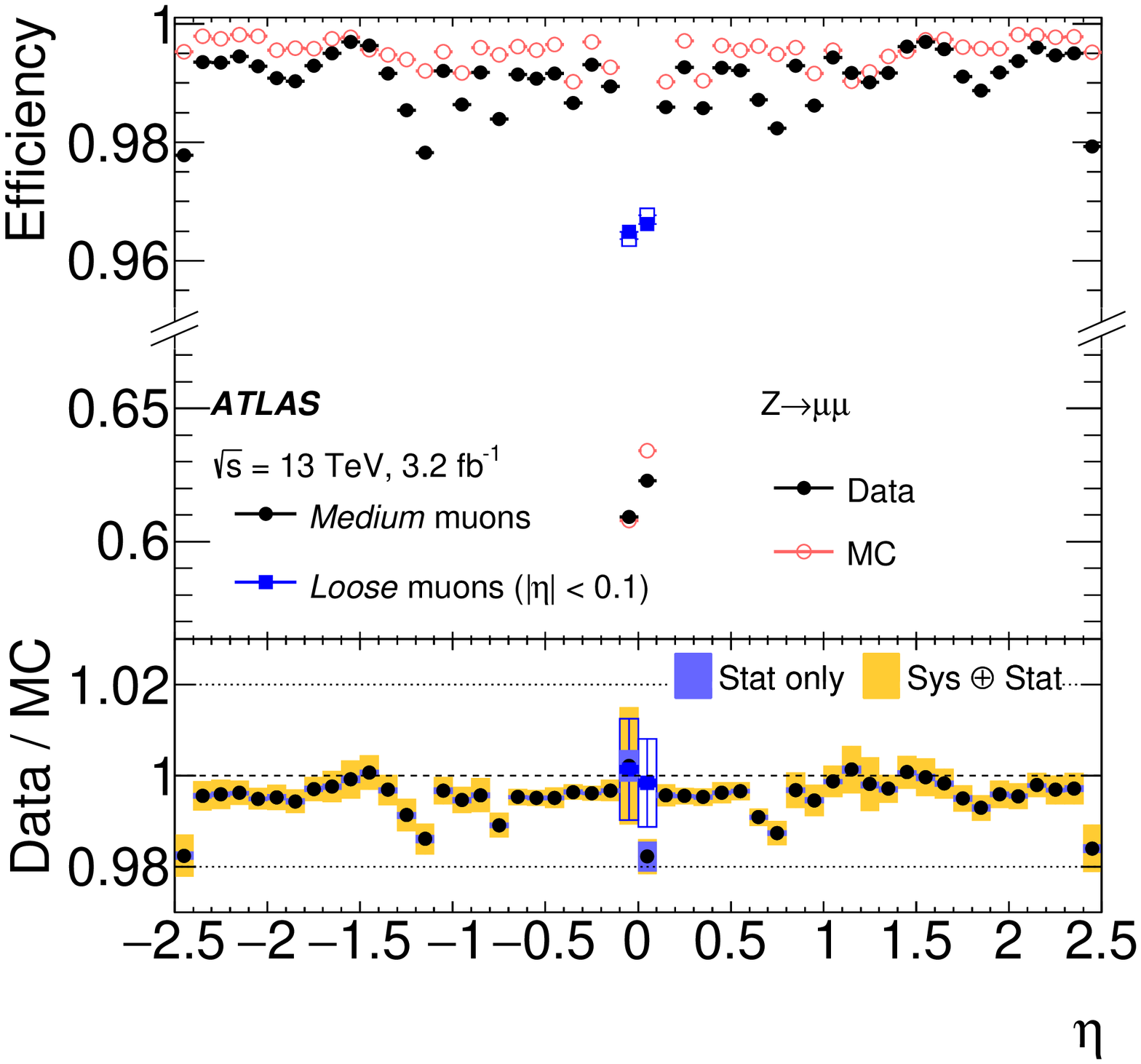}

    \includegraphics[width=0.49\linewidth]{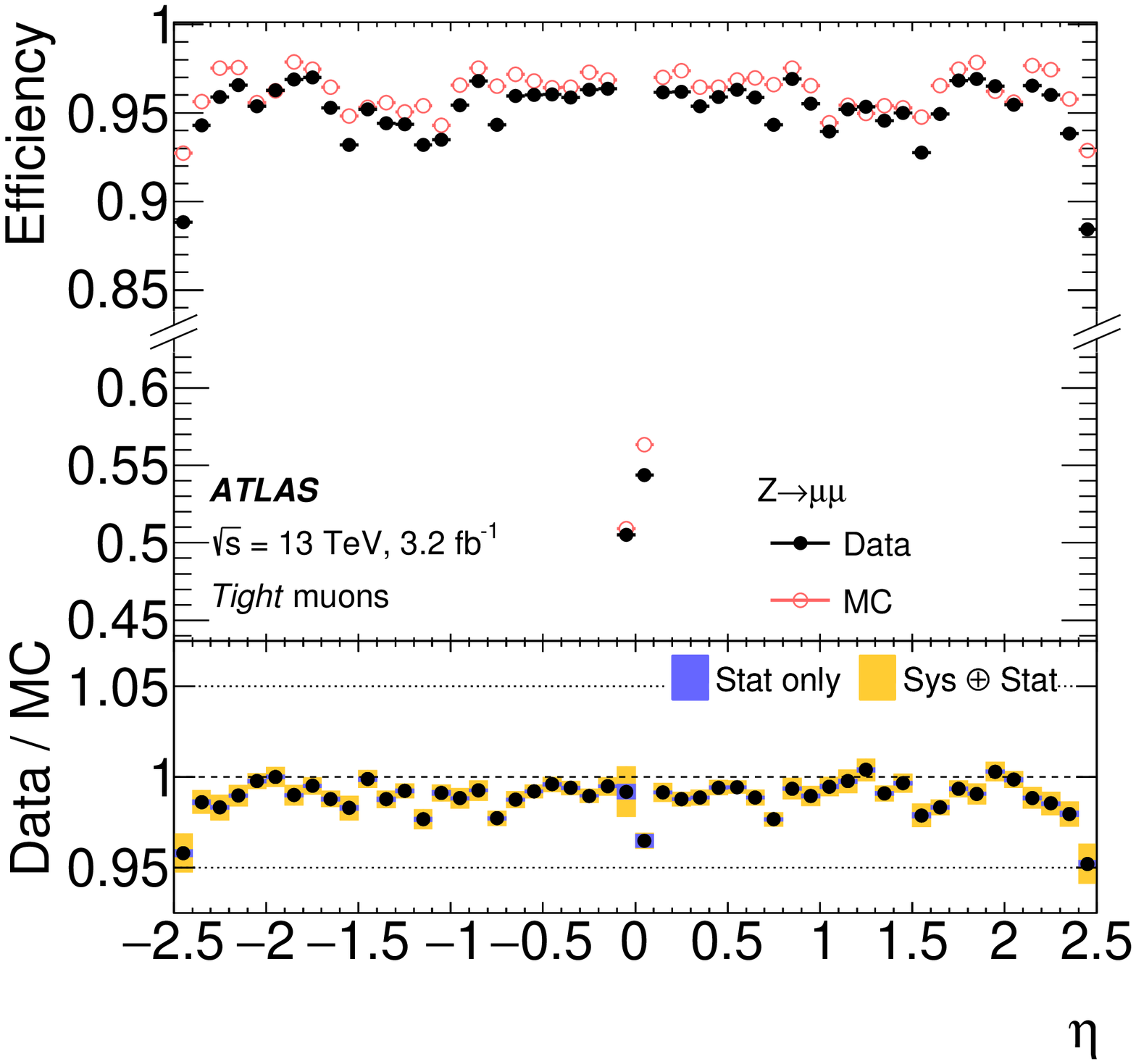}
    \includegraphics[width=0.49\linewidth]{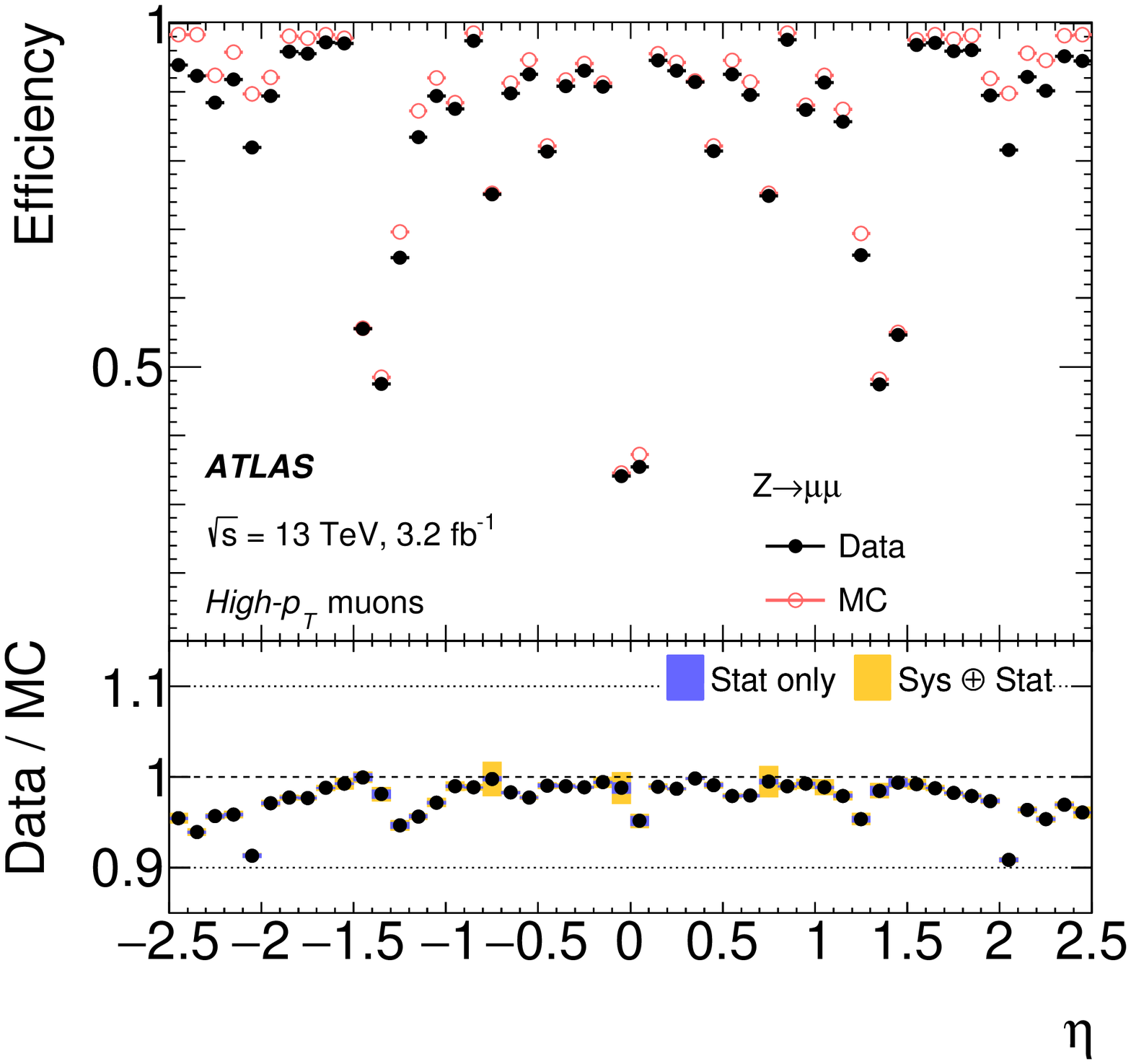}

\caption{Muon reconstruction efficiency as a function of $\eta$ measured in $\Zmm$ events for muons with $\pt>10$~\GeV\ shown for \emph{Medium} (top), \emph{Tight} (bottom left),
and \emph{High-p$_{\rm T}$} (bottom right) muon selections. In addition, the top plot also shows  the efficiency of the \emph{Loose} selection (squares) in the region $|\eta|<0.1$ where the
 \emph{Loose} and  \emph{Medium} selections differ significantly.
  The error bars on the efficiencies indicate the statistical uncertainty.
Panels at the bottom show the ratio of the measured to predicted efficiencies, with statistical and systematic uncertainties.}
\label{Fig:AllTypeEff_Z}
\end{center}
\end{figure}

\FloatBarrier

Figure~\ref{Fig:MediumEff_pt} shows the reconstruction efficiencies for the \emph{Medium} muon selection
as a function of  transverse momentum, including results from $\Zmm$ and $\Jpsimm$,  for muons
with $0.1<|\eta|<2.5$.
The efficiency is stable and slightly above 99\% for $\pt>6$~\GeV.
Values measured from $\Jpsimm$ and $\Zmm$ events are in  agreement in the overlap
region between 10 and 20~\GeV. The efficiency scale factors are also found to be compatible.

\begin{figure}[!h]
\begin{center}
    \includegraphics[width=0.99\linewidth]{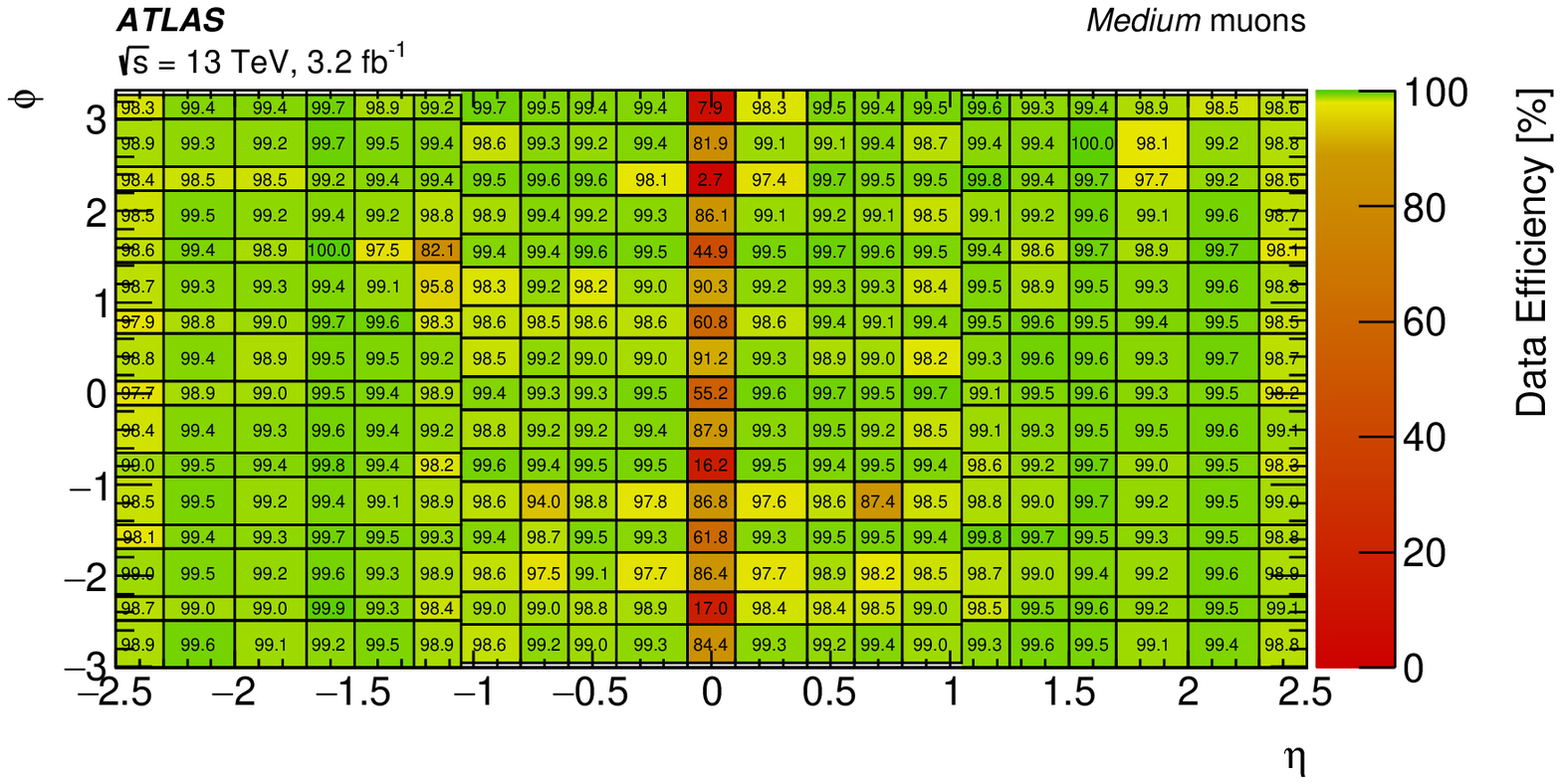}
    \includegraphics[width=0.99\linewidth]{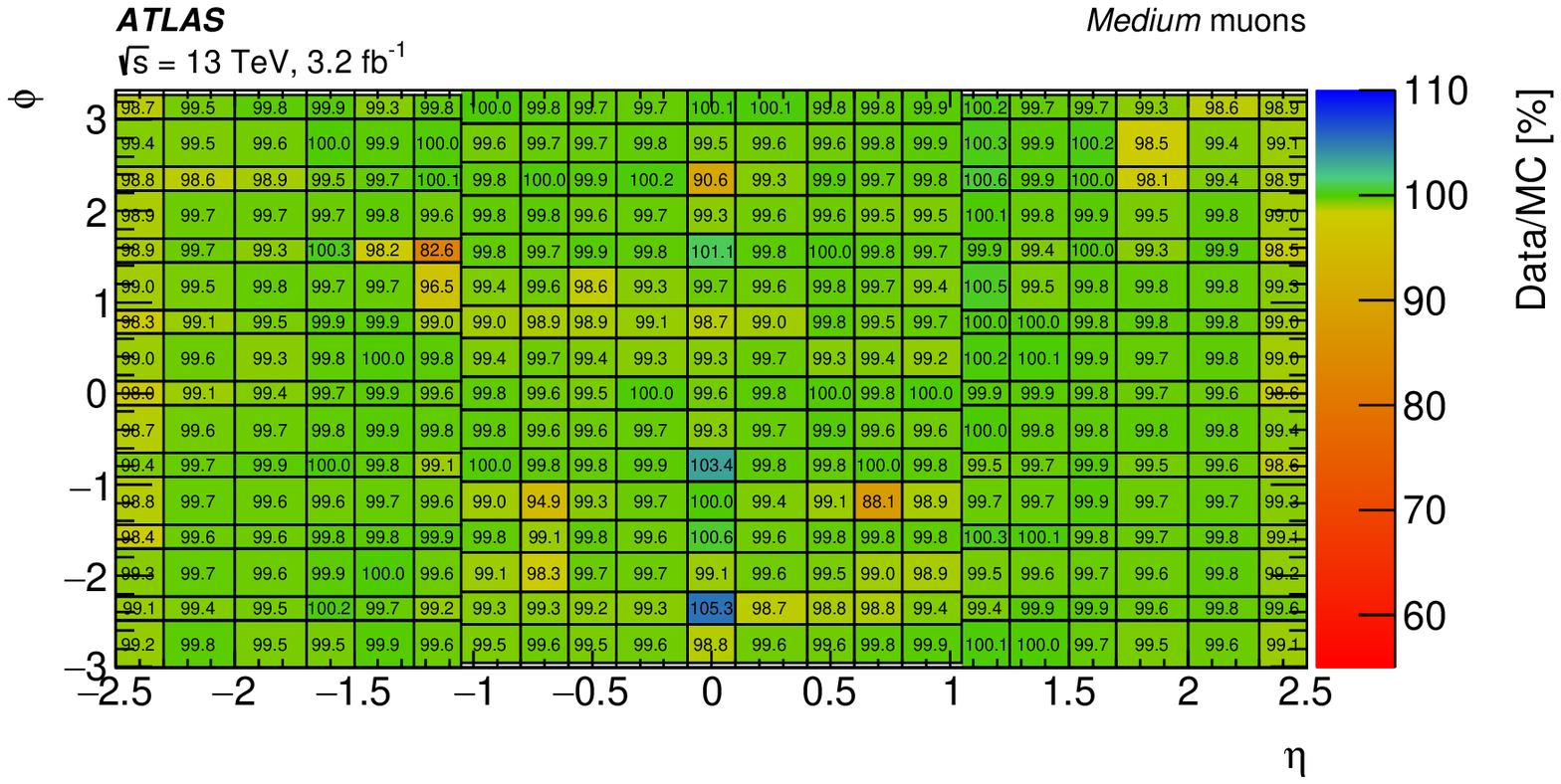}
  \caption{Reconstruction efficiency measured in data (top),
    and the data/MC efficiency scale factor (bottom) for \emph{Medium} muons
    as a function of $\eta$ and $\phi$ for muons with $\pt>10$~\GeV\ in $\Zmm$ events.
    The thin white bins visible in the region $| \phi | \sim \pi$ are due to the different bin boundaries in $\phi$ in the  endcap and barrel regions. }
 \label{Fig:Medium2DEff_Z}
\end{center}
\end{figure}

\begin{figure}[bh]
\begin{center}
    \includegraphics[width=0.49\linewidth]{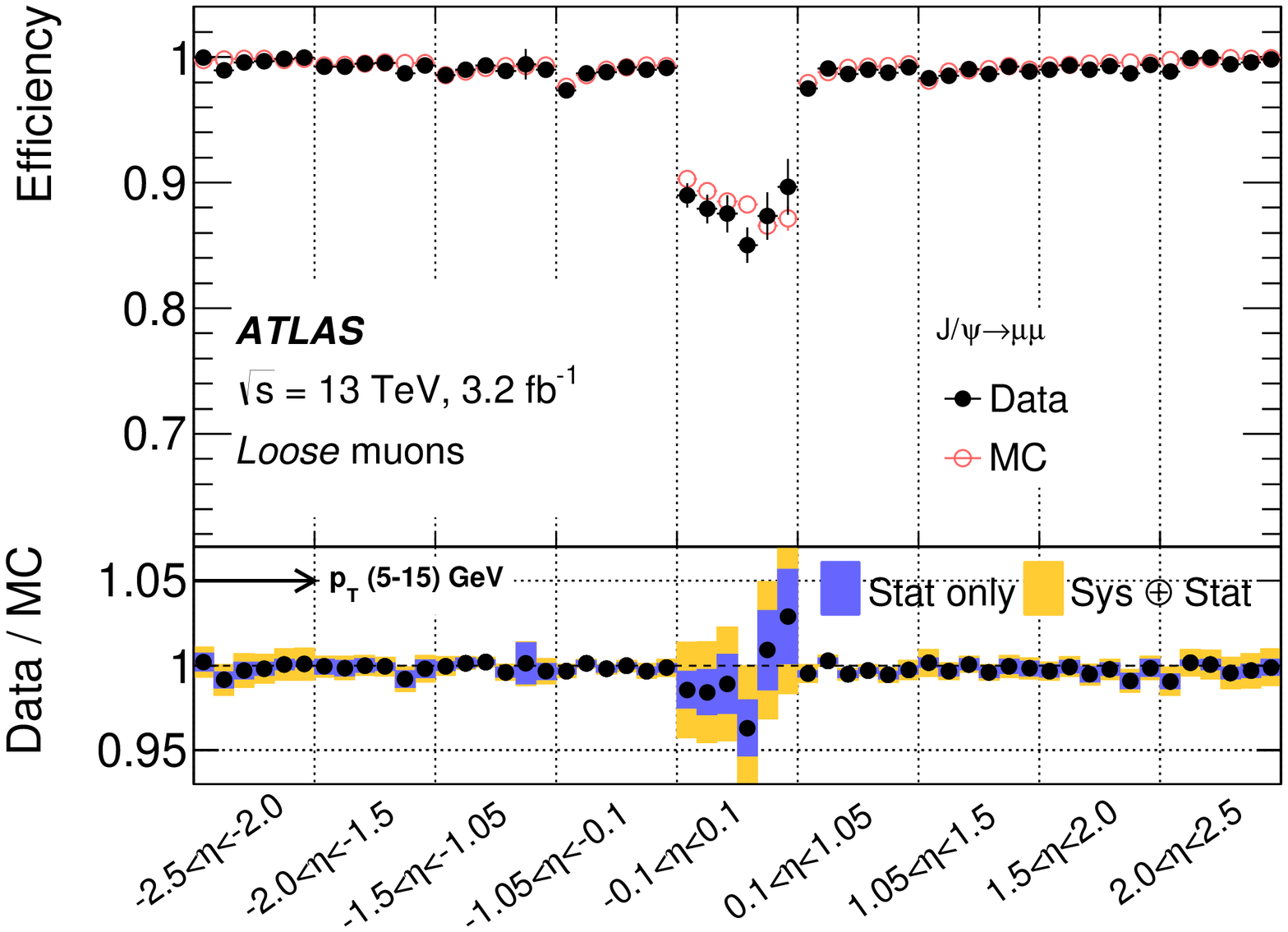}
    \includegraphics[width=0.49\linewidth]{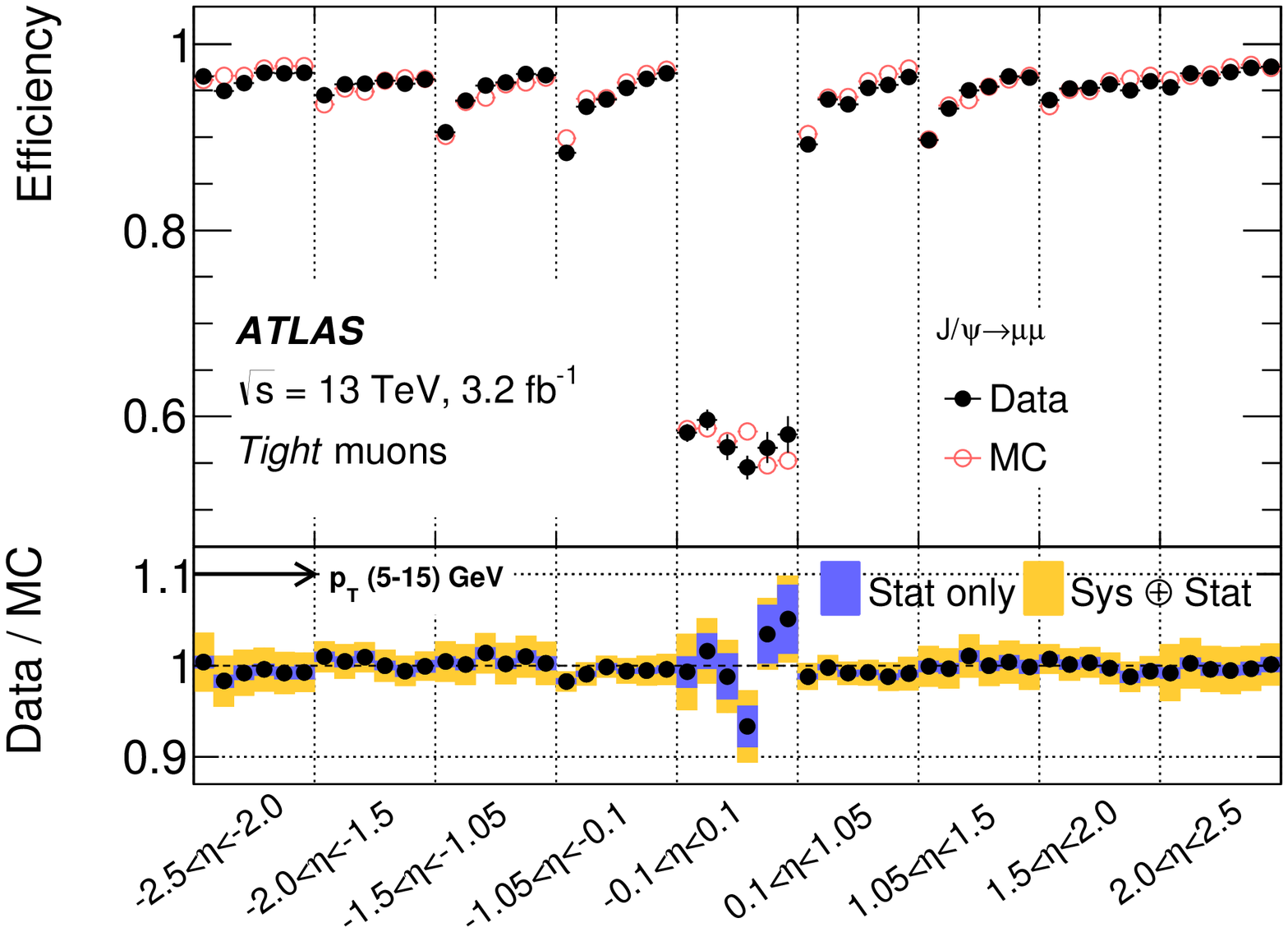}

\caption{Muon reconstruction efficiency  in different $\eta$ regions measured in $\Jpsimm$ events
  for \emph{Loose} (left) and \emph{Tight} (right) muon selections.
  Within each $\eta$ region, the efficiency is measured in six $\pt$ bins
  (5--6, 6--7, 7--8, 8--10, 10--12, and 12--15 GeV).
  The resulting values are plotted  as distinct measurements in each $\eta$ bin with \pt\ increasing from 5 to 15~\GeV\ going from left to right.
  The error bars on the efficiencies indicate the statistical uncertainty.
The panel at the bottom shows the ratio of the measured to
  predicted efficiencies, with statistical and systematic uncertainties.}
\label{Fig:AllTypeEff_JPsi}
\end{center}
\end{figure}

\begin{figure}[!h]
\begin{center}
    \includegraphics[width=0.8\linewidth]{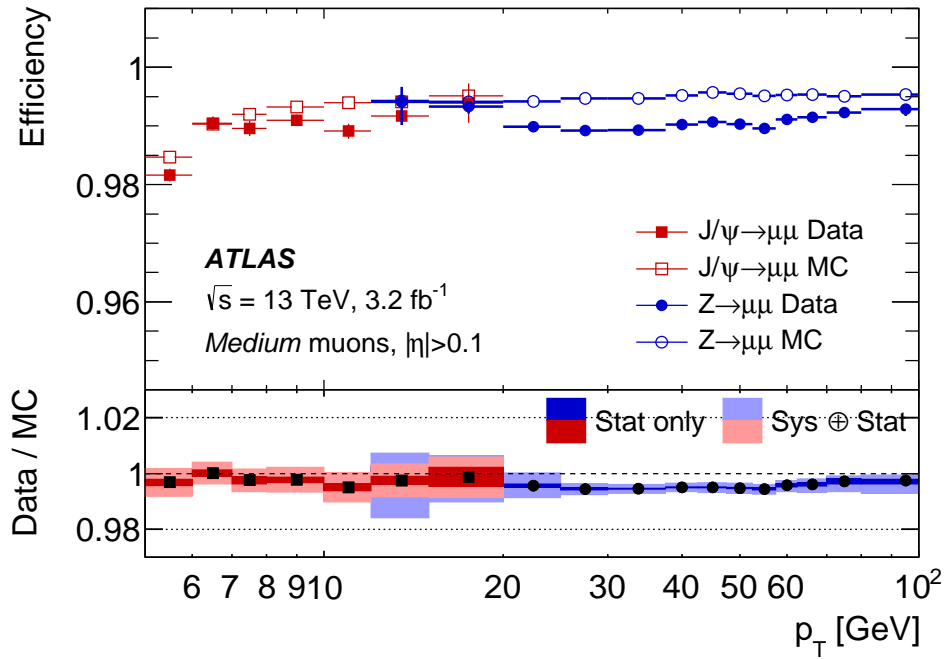}
   \caption{Reconstruction efficiency for the \emph{Medium} muon selection  as a function
     of the $\pt$ of the muon, in the region  $0.1<|\eta|< 2.5$ as obtained with $\Zmm$ and $\Jpsimm$ events.
     The error bars on the efficiencies indicate the statistical uncertainty.
     The panel at the bottom shows the ratio of the measured to predicted efficiencies, with statistical and
     systematic uncertainties.}
   \label{Fig:MediumEff_pt}
\end{center}
\end{figure}

\subsection{Muon reconstruction efficiency for $|\eta|> 2.5$}
\label{sec::muon_reco_forward}

As described in the previous sections, the reconstruction of combined muons is limited
by the ID acceptance to the pseudorapidity region $|\eta|<2.5$.
For $|\eta|>2.5$, the efficiency is recovered by using the ME muons included in the \emph{Loose}
and \emph{Medium} muon selections.
A measurement of the efficiency SF for muons in the region $2.5<|\eta|<2.7$ (high-\eta region) is performed using
the method described in Ref.~\cite{perf2012}.
The number of muons observed in $\Zmm$  decays in the high-\eta region is normalised
to the number of muons observed in the region $2.2<|\eta|<2.5$. This ratio is calculated for both
 data and simulation, applying all known performance corrections to the region $|\eta|<2.5$.
The SFs in the high-\eta region are defined as the ratio of the aforementioned ratios
and are provided in 4 $\eta$ and 16 $\phi$ bins.
The values  of the SFs measured using the 2015 dataset  are close to 0.9 and are determined with a 3--5\% uncertainty.

\section{Isolation}
\label{Sec:Isolation}

Muons originating from the decay of heavy particles, such as $W$, $Z$,  or Higgs bosons,
are often produced isolated from other particles.
Unlike muons from semileptonic decays, which are embedded in jets, these muons
are well separated from other particles in the event.
The measurement of the detector activity around a muon candidate, referred to as {\em muon isolation}, is therefore
a powerful tool for background rejection in many physics analyses.

\subsection{Muon isolation variables }
\label{sec::muon_isol_var}

Two variables are defined to assess muon isolation: a track-based isolation variable
and a calorimeter-based isolation variable.

The track-based isolation variable, $\pt^\mathrm{varcone30}$, is defined as the scalar sum of the transverse momenta of the tracks with $ \pt > $1 \GeV\
in a cone of size $\Delta R = \mathrm{min}\left(10\GeV/\pt^{\mu}, 0.3\right)$ around the muon of transverse momentum $\pt^{\mu}$, excluding the muon track itself.
The cone size is chosen to be $\pt$-dependent to improve the performance
for muons produced in the decay of particles with a large transverse momentum.

The calorimeter-based isolation variable, $\et^\mathrm{topocone20}$, is defined as the sum of the transverse energy of topological clusters~\cite{Aad:2011he} in a cone of size  $\Delta R$ = 0.2 around the muon, after subtracting the contribution from the energy deposit of the muon itself and correcting for pile-up effects. Contributions from pile-up and the underlying event are estimated using the ambient energy-density technique~\cite{Cacciari:2007fd} and are corrected  on an event-by-event basis.

The isolation selection criteria are determined using  the {\em relative isolation variables}, which are defined as the ratio of the track- or  calorimeter-based isolation variables to the transverse momentum of the muon. The distribution of the relative isolation variables in muons from $\Zmm$ events is shown in the top panels of Fig.~\ref{fig:Isol_Var}. Muons included in the plot satisfy the {\it Medium} identification criteria and are well separated from the other muon from the $Z$ boson  (${\Delta{}R}_{\mu\mu}>0.3$). The bottom panel shows the ratio of data to simulation.

\begin{figure*}[!h]
\begin{center}
 \includegraphics[width=0.42\linewidth]{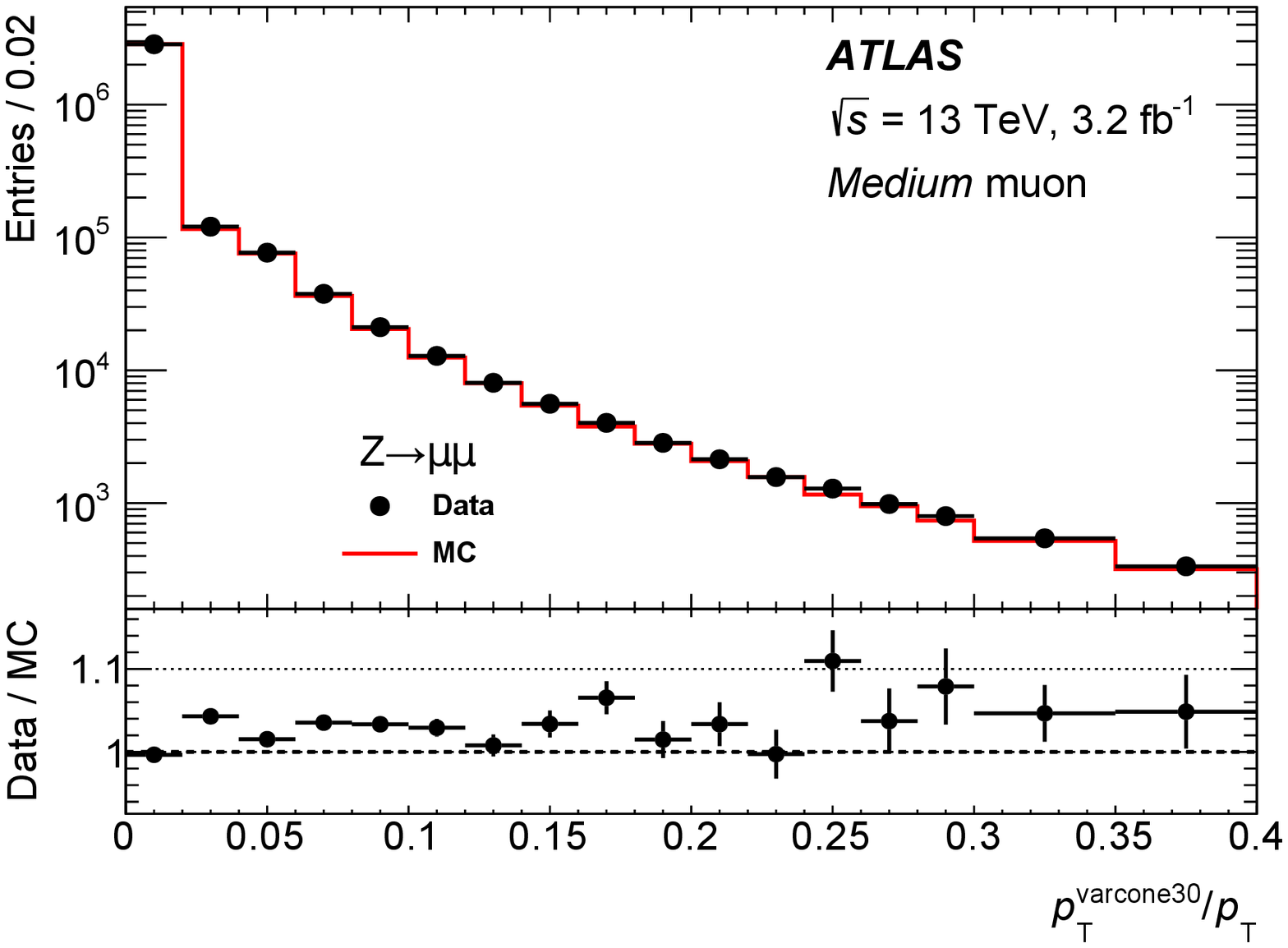}
 \includegraphics[width=0.42\linewidth]{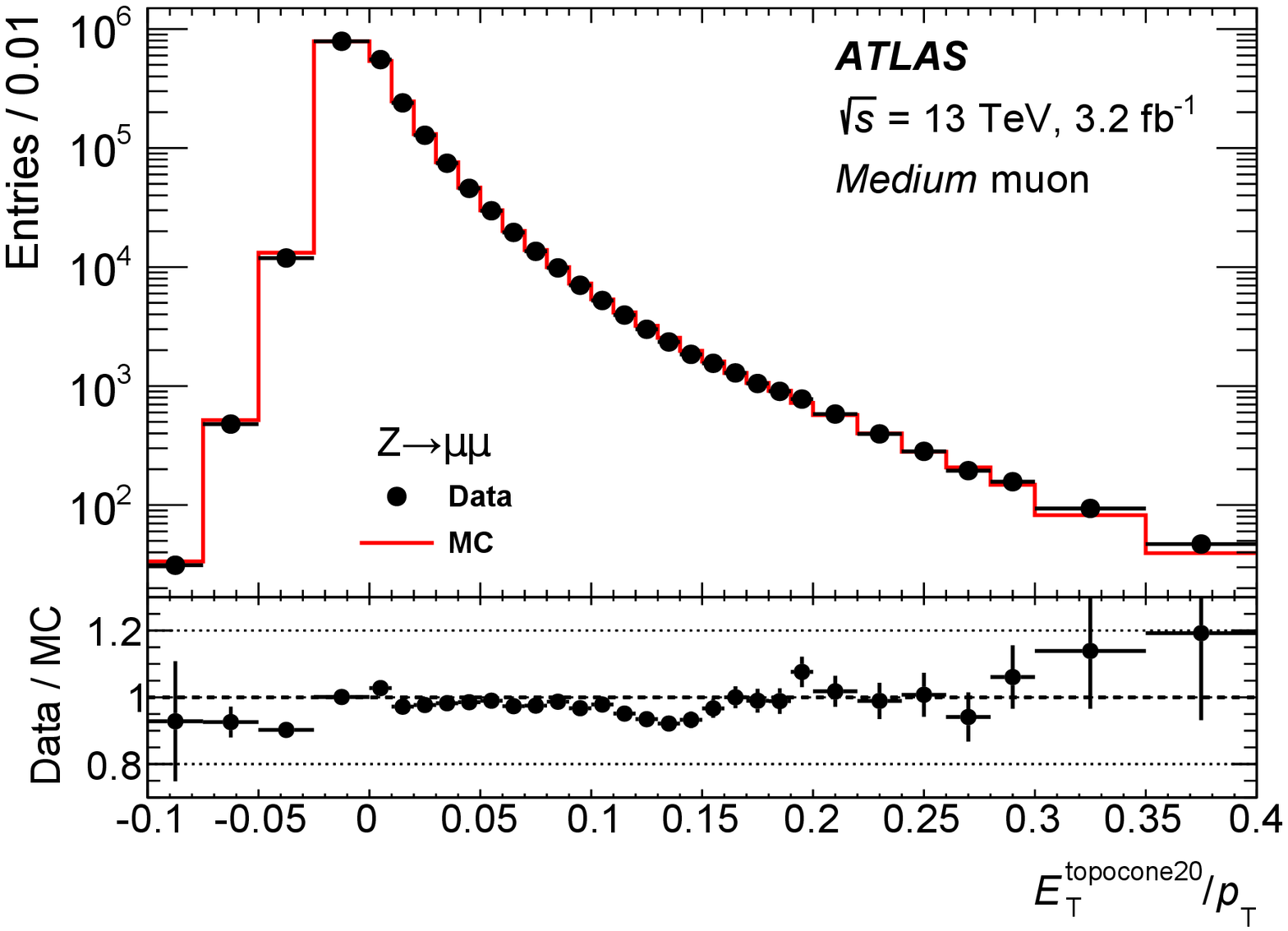}
    \caption{
Distributions of the track-based (left)  and the calorimeter-based (right) relative isolation variables
measured in $Z\rightarrow\mu\mu$ events. Muons are selected by the \emph{Medium} identification algorithm.
The dots show the distribution for data while the  histograms show the distribution from simulation.
The bottom panels show the ratio of data to simulation with the corresponding statistical uncertainty.
The  pile-up reweighted simulated distribution is normalised to the number of events selected in data.
}
\label{fig:Isol_Var}
\end{center}
\end{figure*}

\subsection{Muon isolation performance }
\label{sec::muon_isol_perf}

Seven  isolation selection criteria ({\em isolation working points}) are defined, each optimised
for different physics analyses.  
Table~\ref{Tab:ISoWP} lists  the seven isolation working points with the
discriminating variables and  the criteria used in their definition.
%

\begin{table}
\begin{center}
\renewcommand{\arraystretch}{1.4}
\begin{tabular}{c|c|c}
\hline
\hline
Isolation WP & Discriminating variable(s) & Definition \\
\hline
\emph{LooseTrackOnly} & $\pt^\mathrm{varcone30}/\pt^\mu$                            & 99\% efficiency constant in $\eta$ and \pT \\
\hline
 \emph{Loose} & $\pt^\mathrm{varcone30}/\pt^\mu$, $\et^\mathrm{topocone20}/\pt^\mu$  &  99\% efficiency constant in $\eta$ and \pT \\
\hline
 \emph{Tight} & $\pt^\mathrm{varcone30}/\pt^\mu$, $\et^\mathrm{topocone20}/\pt^\mu$  & 96\% efficiency constant in $\eta$ and \pT \\
\hline
 \emph{Gradient}      & $\pt^\mathrm{varcone30}/\pt^\mu$, $\et^\mathrm{topocone20}/\pt^\mu$  &  $\ge 90 (99)\%$ efficiency at 25 (60) \GeV \\
\hline
 \emph{GradientLoose} & $\pt^\mathrm{varcone30}/\pt^\mu$, $\et^\mathrm{topocone20}/\pt^\mu$ &  $\ge 95 (99)\%$ efficiency at 25 (60) \GeV \\
\hline
 \emph{FixedCutTightTrackOnly} & $\pt^\mathrm{varcone30}/\pt^\mu$ & $\pt^\mathrm{varcone30}/\pt^\mu < 0.06 $\\
\hline
 \emph{FixedCutLoose} & $\pt^\mathrm{varcone30}/\pt^\mu$, $\et^\mathrm{topocone20}/\pt^\mu$  &  $\pt^\mathrm{varcone30}/\pt^\mu<0.15$, $\et^\mathrm{topocone20}/\pt^\mu<0.30$ \\
\hline
\hline
\end{tabular}
\caption{Definition of the seven isolation working points.
The  discriminating variables are listed in the second column
and the criteria used in the definition are reported in the third column.}
\label{Tab:ISoWP}
\end{center}
\end{table}

The efficiencies for the seven isolation working points are measured in data and simulation in  $\Zmm$ decays using the tag-and-probe method described in Section~\ref{Sec:Efficiency}. To avoid probe muons in the vicinity of a jet, the angular separation ${\Delta{}R}$ between the probe muon and the closest jet, reconstructed using an anti-$k_t$ algorithm~\cite{Cacciari:2008gp} with radius parameter 0.4 and with a transverse momentum greater than 20 \GeV, is required to be greater than 0.4.
In addition, the two muons originating from the $Z$ boson decay are required to be separated by ${\Delta{}R}_{\mu\mu}>0.3$.
Figure~\ref{fig:Isol_Perf_Paper} shows the isolation efficiency measured for {\em Medium} muons in data and simulation as a function of the muon \pt\ for the \emph{LooseTrackOnly}, \emph{Loose}, \emph{GradientLoose}, and \emph{FixedCutLoose}  working points,
with the respective data/MC ratios included in the bottom panel.
The systematic uncertainties in the SFs are estimated
by varying the background contributions within their uncertainties and
by varying some of the selection criteria, such as the invariant mass selection window, the isolation of the tag muon, the minimum quality of the probe muon,
the opening angle between the two muons, and the $\Delta R$ between the probe muon and the closest jet.
In Fig.~\ref{fig:Isol_Perf_Paper},  the largest systematic uncertainty contribution over the entire \pT\ region arises from having neglected the $\eta$ dependence of the SFs, which are usually provided as a function of $\eta$ and \pt .
In the low-\pT\ region, other important contributions are due to the background estimation and the mass window variation,
while the high-\pT\ region is dominated by statistical uncertainties in data and simulation.
The total uncertainty is at the per mille level over a wide range of \pT\ and reaches the percent level in the high-\pT\ region.
The suppression factor for  muons from light mesons or $b$/$c$ semileptonic decays is estimated from simulation and depends on the isolation working point,
ranging from a minimum of 15 for {\em LooseTrackOnly} to  a maximum of 40 for {\em Gradient}.

\begin{figure*}[!h]
\begin{center}
 \includegraphics[width=0.4\linewidth]{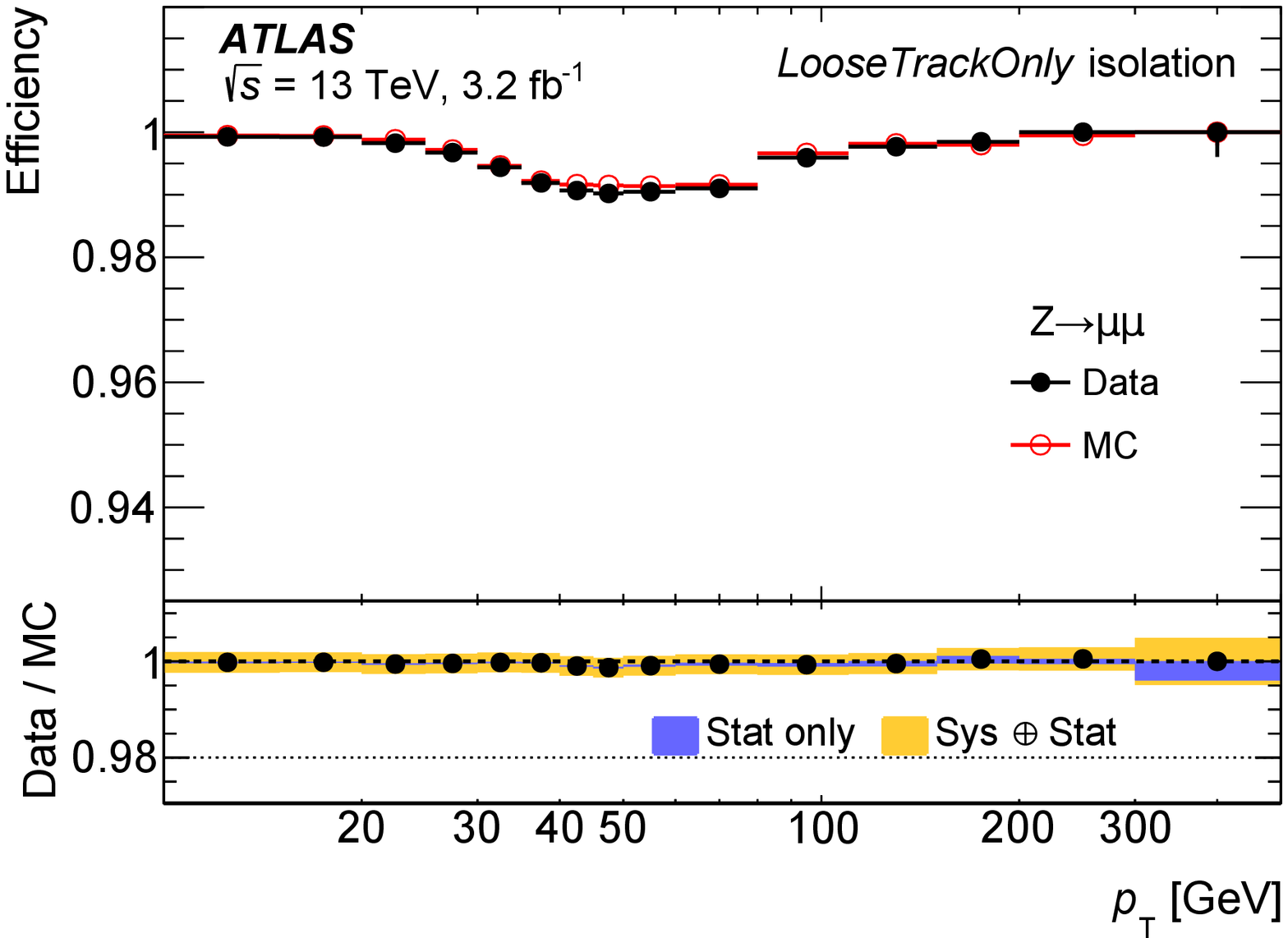}
 \includegraphics[width=0.4\linewidth]{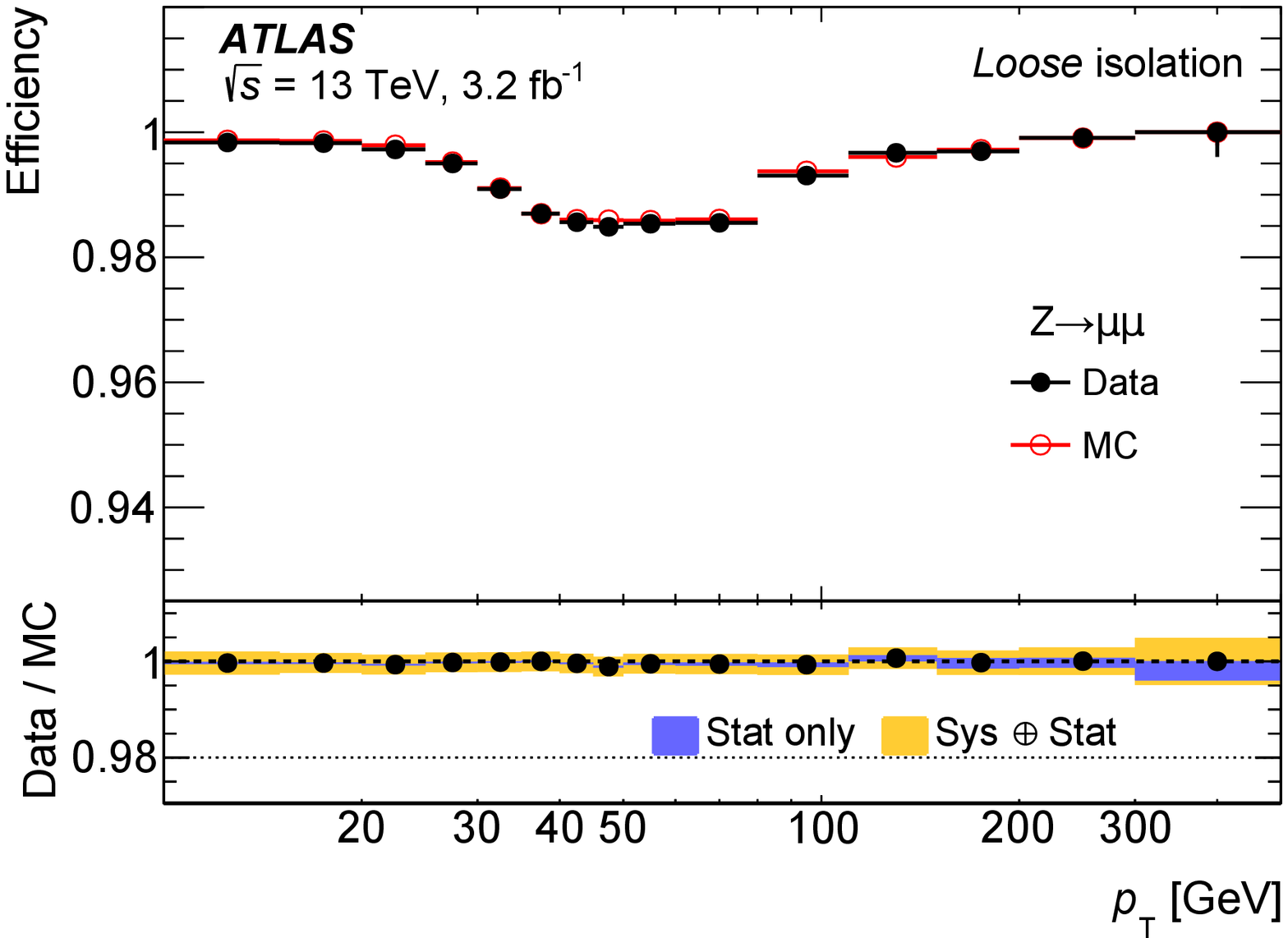}
 \includegraphics[width=0.4\linewidth]{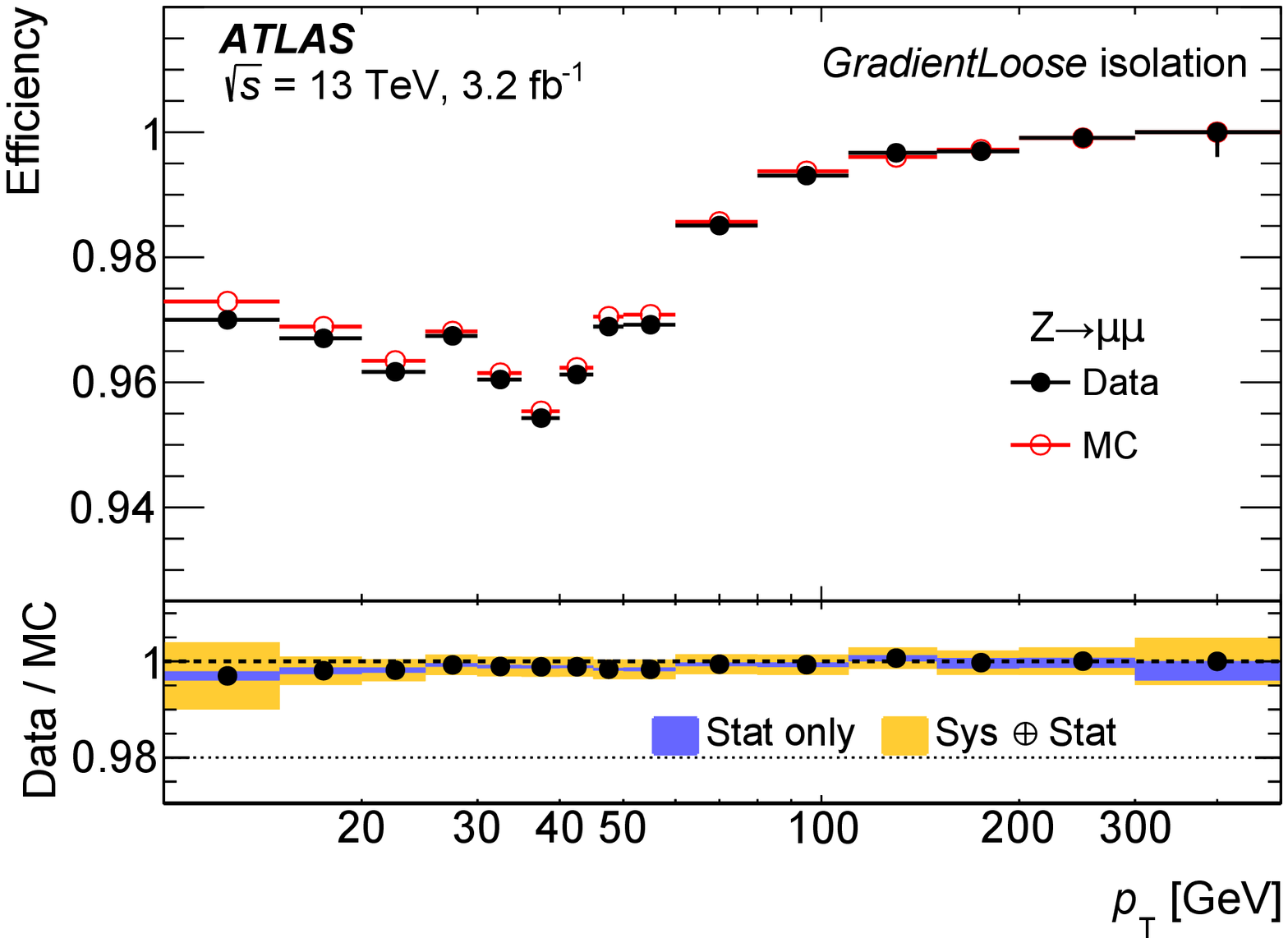}
 \includegraphics[width=0.4\linewidth]{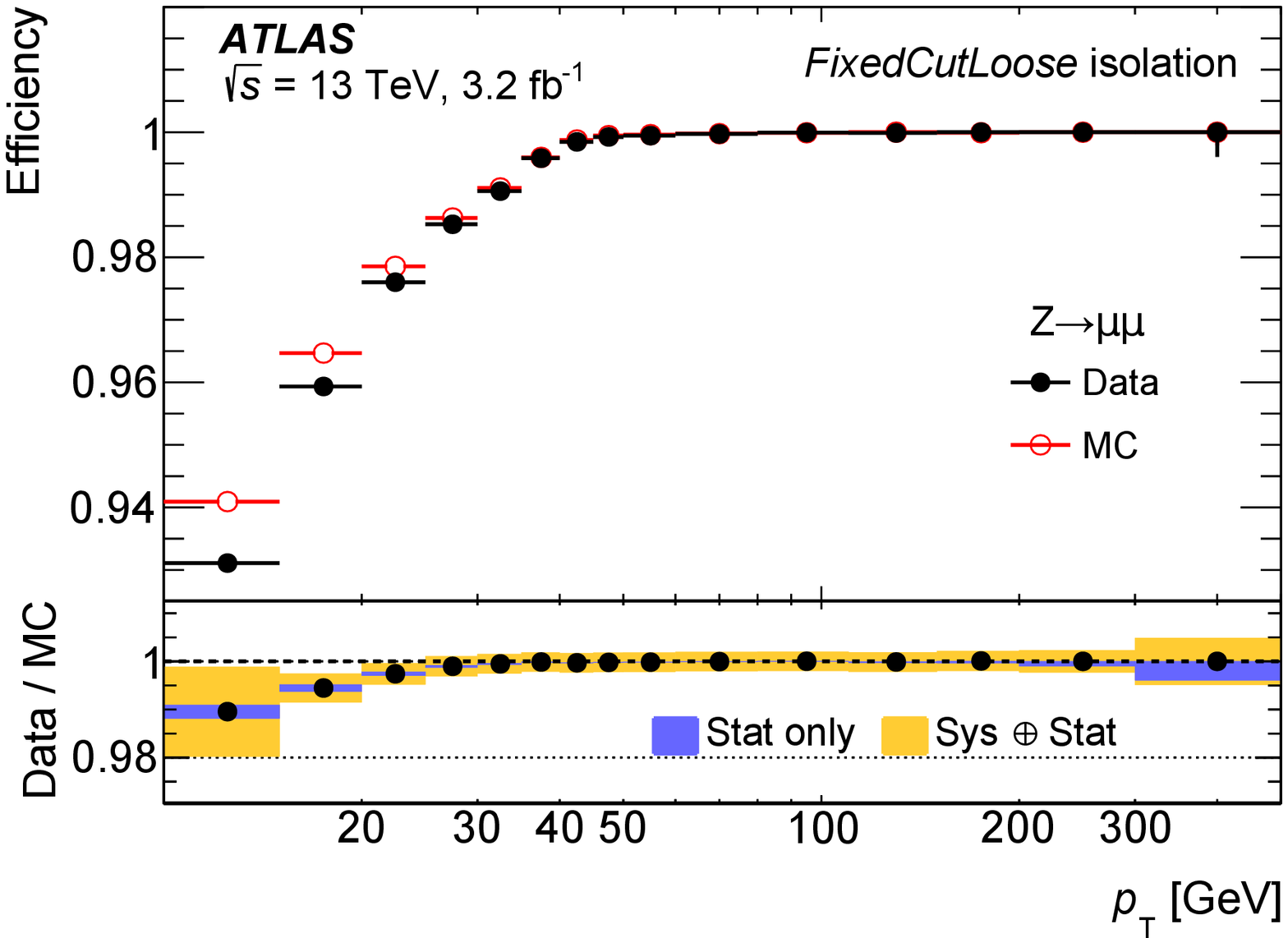}
 \caption{Isolation efficiency for the
   \emph{LooseTrackOnly} (top left), \emph{Loose} (top right), \emph{GradientLoose} (bottom left), and \emph{FixedCutLoose} (bottom right)
   muon isolation working points.
   The efficiency is shown  as a function of the muon transverse momentum \pt\ and is  measured in $Z\rightarrow\mu\mu$ events.
   The full (empty) markers indicate the efficiency measured in data (MC) samples. The errors shown on the efficiency are statistical only.
   The bottom panel shows the ratio of the efficiency measured in data and simulation, as well as the statistical uncertainties
   and combination of statistical and systematic uncertainties.}
\label{fig:Isol_Perf_Paper}
\end{center}
\end{figure*}

\section{Momentum scale and resolution}
\label{Sec:Resolution}
The muon momentum scale and resolution are studied using
 $\Jpsimm$  and $\Zmm$ decays.
Although the simulation contains an accurate description of the ATLAS detector,
the level of detail is not enough to describe the muon momentum scale to the per mille level and the muon momentum resolution to the percent level.
To obtain such of agreement between data and simulation, a set of corrections is applied to the simulated muon  momentum.
The methodology used to extract these corrections is described in
\refS{Sec:CorrectionExtraction}.
In \refS{sec::validation}, measurements of the muon momentum scale and resolution in data and simulation
are presented for various detector  regions and for a wide range of \pT .
To improve the precision of the procedure,
the $\pT$ and $\eta$ distributions  of the $Z$ and $J/\psi$ resonances in simulation are reweighted to the distributions observed in data.

\subsection{Muon momentum calibration procedure}\label{Sec:CorrectionExtraction}

In the following, the ``muon momentum calibration'' is defined as the procedure used to identify
the corrections to the simulated muon transverse momenta reconstructed in the ID and MS sub\-de\-tec\-tors
to precisely describe the measurement of the same quantities in data.
Only CB muons are used to extract the calibration parameters.
The transverse momentum of the ID and MS components of a CB track, referred to as $\pt^{\textrm{ID}}$ and $\pt^{\textrm{MS}}$, respectively,    are used.
The ID (MS) tracks are reconstructed using the hits from the ID (MS) detector and are extrapolated to the interaction point.
In the case of MS tracks,
the fit corrects for the energy loss in the calorimeters as described earlier.

The corrected transverse momentum, $\pt^{\rm Cor,Det}$  (Det$ = {\rm ID, MS}$), is described by  the following equation:
\begin{eqnarray}\label{eq:pt_cor}
  \pt^{\rm Cor,Det} &=&  \frac{\pt^{\rm MC,Det} + \sum\limits^{1}_{n=0} s^{\rm Det}_{n}(\eta,\phi)\left(\pt^{\rm MC,Det}\right)^{n}}{1+\sum\limits^{2}_{m=0} \Delta r^{\rm Det}_{m}(\eta,\phi)\left(\pt^{\rm MC,Det}\right)^{m-1}g_{m}},
\end{eqnarray}
where
 $\pt^{\rm MC, Det}$ is the uncorrected transverse momentum in simulation,
 $g_{m}$ are normally distributed random variables
with zero mean and unit width,
and the terms  $\Delta r_{m}^{\rm Det}(\eta,\phi)$ and $s_n^{\rm Det}(\eta, \phi)$  describe the momentum resolution smearing and the scale corrections applied in a specific ($\eta$, $\phi$) detector region, respectively.

The corrections described in Eq.~(\ref{eq:pt_cor}) are defined in $\eta$--$\phi$ detector regions that are homogeneous
in terms of detector technology and  performance.
Both the ID and the MS are  divided into  18 pseudorapidity regions.
In addition, the MS is divided into two $\phi$ bins separating the two types of $\phi$ sectors: those
that include the magnet coils ({\em small sectors}) and those
 between two coils ({\em large sectors}).
The  small and large MS sectors employ independent alignment techniques and cover detector areas with different material distribution.
Therefore, relevant scale and resolution differences exist.

The numerator of Eq.~(\ref{eq:pt_cor}) describes the momentum scales.
The $s_1^{\rm Det} $ term corrects for inaccuracy in the description of the magnetic field integral and the dimension of the detector in the direction perpendicular to the magnetic field.
The $s^{\rm MS}_0(\eta, \phi)$ term models the effect on the MS momentum from the inaccuracy in the  simulation of the energy loss
in the calorimeter and other materials between the interaction point and the MS.
As the energy loss between the interaction point and the ID is negligible,  $s^{\rm ID}_0(\eta)$ is set to zero.

The denominator of Eq.~(\ref{eq:pt_cor}) describes the momentum smearing that broadens the relative \pt\ resolution in simulation,
 $\sigma(\pt)/ \pt$,
to properly describe the data.
The corrections to the resolution assume that the relative \pt\ resolution can be parameterized as follows:
\begin{equation}\label{eq:qoverpt_reso}
 \frac{ \sigma(\pt)}{\pt} = r_0/\pt \oplus r_1 \oplus r_2\cdot \pt\textrm{,}
\end{equation}

with $\oplus$ denoting a sum in quadrature. In  Eq.~(\ref{eq:qoverpt_reso}), the first term
 accounts mainly for fluctuations of the energy loss in the traversed material,
the second term accounts mainly for multiple scattering, local magnetic field inhomogeneities
and local radial displacements of the hits,  and the third term mainly describes intrinsic resolution effects caused by the spatial resolution of the hit measurements and by residual misalignment of the muon spectrometer.
The energy loss term is negligible in both the ID and MS measurements, and therefore $ \Delta r^{\rm ID}_{0}$ and $ \Delta r^{\rm MS}_{0}$ are set to zero.

The corrected momentum of the combined muons, $\pt^{\rm Cor,CB}$,  is obtained by combining the ID and MS corrected momenta
 using a weighted average:
  \begin{equation}\label{eq:cb_correction1}
   \pt^{\rm Cor,CB} = f\cdot \pt^{\rm Cor,ID} + (1 - f)\cdot \pt^{\rm Cor, MS}\textrm{,}
 \end{equation}
with the weight $f$ derived from the following linear equation
\begin{equation}\label{eq:cb_correction2}
  \pt^{\rm MC,CB} = f\cdot \pt^{\rm MC,ID} + (1 - f)\cdot \pt^{\rm MC, MS}
\end{equation}
which assumes that the relative contribution of the two subdetectors to the combined track remains unchanged before and after momentum corrections.

\subsubsection{Determination of the \pt\ calibration constants}
\label{sec:corr_extract}

The MS and ID correction parameters contained in Eq.~(\ref{eq:pt_cor})
are extracted from data
using a binned maximum-likelihood fit with templates derived from simulation
which compares the invariant mass distributions for $\Jpsimm$ and $\Zmm$ candidates in data and simulation.
The  exceptions are $ \Delta r^{\rm ID}_{0}$, $ \Delta r^{\rm MS}_{0}$, and  $s^{\rm ID}_0$,
which  are set to zero,
and  $ \Delta r^{\rm MS}_{2}$, which  is determined from alignment studies using
special runs with the toroidal magnetic field off.

The $\Jpsimm$ and $\Zmm$ candidates are selected by requiring two oppositely charged CB muons satisfying the {\em Medium} identification criteria.
Both muons must have impact parameters compatible with tracks produced by the primary interaction and pseudorapidity within the
acceptance of both the ID and MS detectors ($|\eta|<$2.5).
Both muons from $\Jpsimm$ ($\Zmm$) candidate decays are required to have momenta in the range 5--20 (22--300)~\GeV\
and to form an invariant mass in the range 2.65--3.6 (76--106) \GeV.
Muons from   $Z$ boson decays need to be isolated, while no isolation criterion is  imposed on muons from $J/\psi$ decays.

The extraction of the correction parameters is performed in $\eta$--$\phi$ regions of fit (ROFs) defined separately for the ID and the MS.
Events are assigned to a specific ROF if at least one muon falls in the corresponding $\eta$--$\phi$ region.

The ID corrections are extracted using the distributions of the ID dimuon invariant mass, $m_{\mu\mu}^{\textrm{ID}}$.
To enhance the sensitivity to  $\pt$-dependent correction effects, the  $m_{\mu\mu}^{\textrm{ID}}$ is classified according to the $\pt$ of the muons.
For $\Jpsimm$ ($\Zmm$) decays, the fit is performed
in two exclusive categories defined requiring the candidates to have  $\pt^{\textrm{ID}}$  of the sub-leading (leading) muon greater than 5 or 9 (22 or 47) \GeV, respectively.

Similarly, the MS corrections are extracted using the distributions of the MS-reconstructed dimuon invariant mass, $m_{\mu\mu}^{\textrm{MS}}$.
Since there are more  parameters and more ROFs in the MS version of Eq.~(\ref{eq:pt_cor}),
an additional variable is added to the MS fit.
This is  defined by the following equation
\begin{equation}
  \label{eq:rho}
  \rho= \frac{\pt^{\textrm{MS}} - \pt^{\textrm{Cor,ID}}}{\pt^{\textrm{Cor,ID}}},
\end{equation}
which represents  the $\pt$ imbalance between the measurement in the ID and in the MS.
In Eq.~(\ref{eq:rho}), the momentum  of the ID, $p_{\rm T}^{\mathrm{Cor,ID}}$,  contains the appropriate \pt\ corrections.
The  variable $\rho$ is  used only in $\Zmm$ candidate events and is
binned according to $\pt^{\textrm{MS}}$ of the muon
with lower bin boundaries  of $\pt^{\textrm{MS}}={22, 35, 47, 60, 90}$~\GeV.

Templates for the  $m_{\mu\mu}^{\textrm{ID}}$, $m_{\mu\mu}^{\textrm{MS}}$, and $\rho$ are built using $\Jpsimm$ and  $\Zmm$ simulated signal  samples.
In the $\Zmm$ sample, the small background component (approximately 0.1\%)  is extracted from simulation and added to the templates.
A much larger  (about 15\%) non-resonant background from decays of light and  heavy hadrons and from continuum Drell--Yan production is present in the $\Jpsimm$ sample.
As this background is not easy to simulate,  a data-driven approach is used. The dimuon invariant mass distribution in data is fitted in each ROF using
a Crystal Ball function  added to an exponential background distribution in the ID and MS fits.
The background model and its normalisation are then used in the template fit.

The results are shown in Tables~\ref{tab:id_cor} and~\ref{tab:ms_cor}, averaged over three $\eta$ regions.
The quoted errors include systematic uncertainties
evaluated by varying several parameters of the template fit.
The main contributions to the final systematic uncertainty are:
\begin{itemize}
  \item Mass window width for the $\Zmm$ candidate selection. Non-Gaussian smearing effects are accounted for
by varying the $m_{\mu\mu}$ selection by $\pm 5$~\GeV.
  \item Background parameterization for the $J/\psi$ fit as well as increased muon $\pt$ cut (from 5 to 7 \GeV) to reduce the  weight of the contribution of low-$\pt$ muons.
  \item Scale parameter for the ID corrections obtained by fitting separately the $\Jpsimm$ and $\Zmm$ samples, to include possible non-linear scale effects.
  \item As $\Delta r_2^{\mathrm{MS}}$ is sensitive to the alignment of the MS chambers, its systematic uncertainty is determined from alignment studies performed on special runs where the toroidal magnetic field was turned off.
\end{itemize}

\begin{table*}[!thb]
  \centering
\begin{tabular}{c r@{}l r@{}l r@{}l}
\toprule
Region & \multicolumn{2}{c}{$\Delta r_1^{\textrm{ID}}(\times 10^{-3})$} & \multicolumn{2}{c}{$\Delta r_2^{\textrm{ID}}$ [TeV$^{-1}$]} & \multicolumn{2}{c}{$s_1^{\textrm{ID}}(\times 10^{-3})$}\\
\midrule
$|\eta|<1.05$         & $4.1$& $\phantom{}^{+0.6}_{-0.9}$ & $0.17$ & $\phantom{}^{+0.04}_{-0.03} $ & $-0.6$ & $\phantom{}^{+0.1}_{-0.2} $  \\[2mm]
$1.05\le |\eta|<2.0$  & $5.5$& $\phantom{}^{+2.5}_{-0.8}$ & $0.34$ & $\phantom{}^{+0.07}_{-0.09} $ & $-0.5$& $\phantom{}^{+0.2}_{-0.5} $  \\[2mm]
$|\eta|\ge 2.0$       & $9$& $\phantom{}^{+9}_{-2}$      & $0.05$ & $\pm{0.01} $       & $1.0$ & $\phantom{}^{+3.5}_{-1.6} $  \\
\bottomrule
\end{tabular}
  \caption{Summary of ID muon momentum resolution and scale corrections used in Eq.~(\ref{eq:pt_cor}),
averaged over three main detector regions.
The corrections are derived in 18 pseudorapidity regions, as described in \refS{Sec:Resolution},
and averaged, assigning a weight to each region proportional to its  $\eta$ width.
The uncertainties represent the sum in quadrature of the statistical and systematic uncertainties.
}
  \label{tab:id_cor}
\end{table*}

\begin{table*}[!thb]
  \centering
\begin{tabular}{cr@{}l r@{}l r@{}l r@{}l}
\toprule
Region & \multicolumn{2}{c}{$\Delta r_1^{\textrm{MS}} (\times 10^{-3})$} & \multicolumn{2}{c}{$\Delta r_2^{\textrm{MS}}$ [TeV$^{-1}$]}  & \multicolumn{2}{c}{$s_0^{\textrm{MS}}$ [MeV]} & \multicolumn{2}{c}{$s_1^{\textrm{MS}}  (\times 10^{-3})$}\\
\midrule
$|\eta|<1.05$ (small)          & $17$& $\pm{1}$    & $0.080$& $\pm{0.006}$   & $-23$& $\pm{5}$          & $-0.9$& $\pm{0.3}   $   \\[2mm]
$|\eta|<1.05$ (large)          & $15$& $\pm{1}$    & $0.162$& $\pm{0.007}$   & $-26$& $\phantom{}^{+8}_{-5}$       & $ 1.8$& $\phantom{}^{+0.4}_{-0.3}$  \\[2mm]
$1.05\le |\eta|<2.0$  (small)  & $25$& $\phantom{}^{+3}_{-1}$ & $0.20$& $\pm{0.03}$     & $-13$& $\pm{6}$          & $-1.4$& $\pm{0.4}   $   \\[2mm]
$1.05\le |\eta|<2.0$  (large)  & $23$& $\phantom{}^{+3}_{-1}$ & $0.160$& $\pm{0.015}$   & $-15$& $\pm{10}$         & $-1.1$& $\phantom{}^{+0.5}_{-0.6}$  \\[2mm]
$|\eta|\ge 2.0$ (small)        & $17$& $\phantom{}^{+3}_{-1}$ & $0.08$& $\pm{0.01}$     & $-6$& $\phantom{}^{+6}_{-7}$        & $ 0.7$& $\phantom{}^{+0.4}_{-0.3}$  \\[2mm]
$|\eta|\ge 2.0$ (large)        & $15$& $\phantom{}^{+4}_{-3}$ & $0.112$& $\pm{0.010}$   & $-3$& $\phantom{}^{+13}_{-10}$      & $ 0.3$& $\phantom{}^{+0.6}_{-0.7}$   \\
\bottomrule
\end{tabular}
\caption{Summary of MS momentum resolution and scale corrections for small and large MS sectors, averaged over three main detector regions.
 The corrections for large and small MS sectors are derived in 18 pseudorapidity  regions, as described in \refS{Sec:Resolution},
and averaged assigning a weight to each region proportional to its  $\eta$ width.
The energy loss term $\Delta r_0^{\textrm{MS}}$ is negligible and therefore fixed to zero in the fit for all $\eta$.
The uncertainties represent the sum in quadrature of the statistical and systematic uncertainties.
}
  \label{tab:ms_cor}
\end{table*}

 \subsection{Dimuon mass scale and resolution after applying momentum corrections}\label{sec::validation}

The  samples of $\Jpsimm$  and $\Zmm$ decays are used to study the muon momentum scales and resolution in data and simulation
and to validate the momentum corrections obtained with the template fit method described in the previous section.

The invariant mass distributions for the $\Jpsimm$ and $\Zmm$ candidates are shown in Fig.~\ref{fig:Calib_InvMass} and compared with uncorrected and corrected simulation.
In the uncorrected simulation,
it is noticeable that  the signal distributions are narrower  and  slightly shifted with respect to data.
After correction, the lineshapes of the two resonances in simulation agree with the data within the systematic uncertainties,
demonstrating the overall effectiveness of the \pt\ calibration.


 \begin{figure*}[th]
   \centering
   \includegraphics[width=0.42\linewidth]{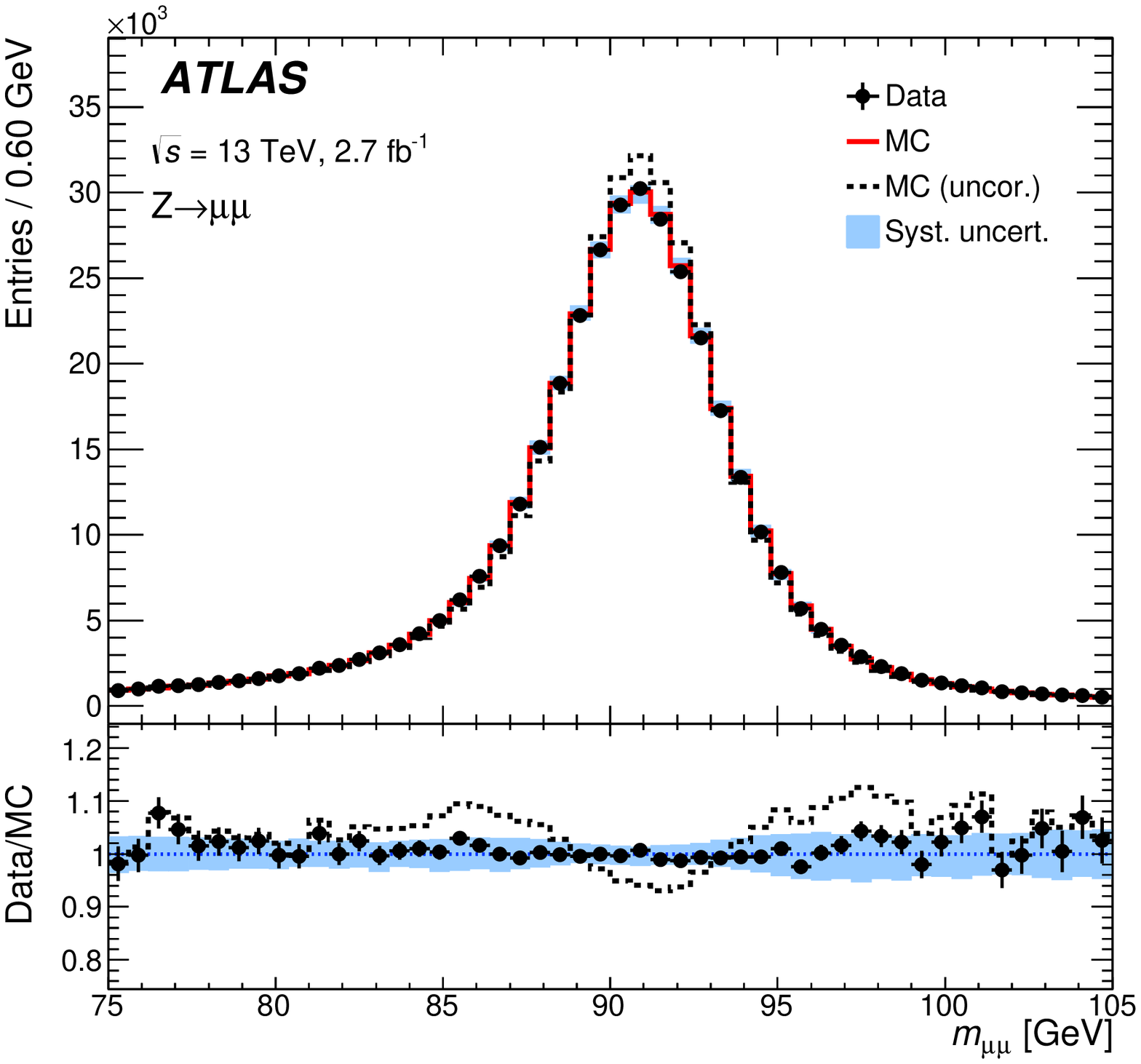}
   \includegraphics[width=0.42\linewidth]{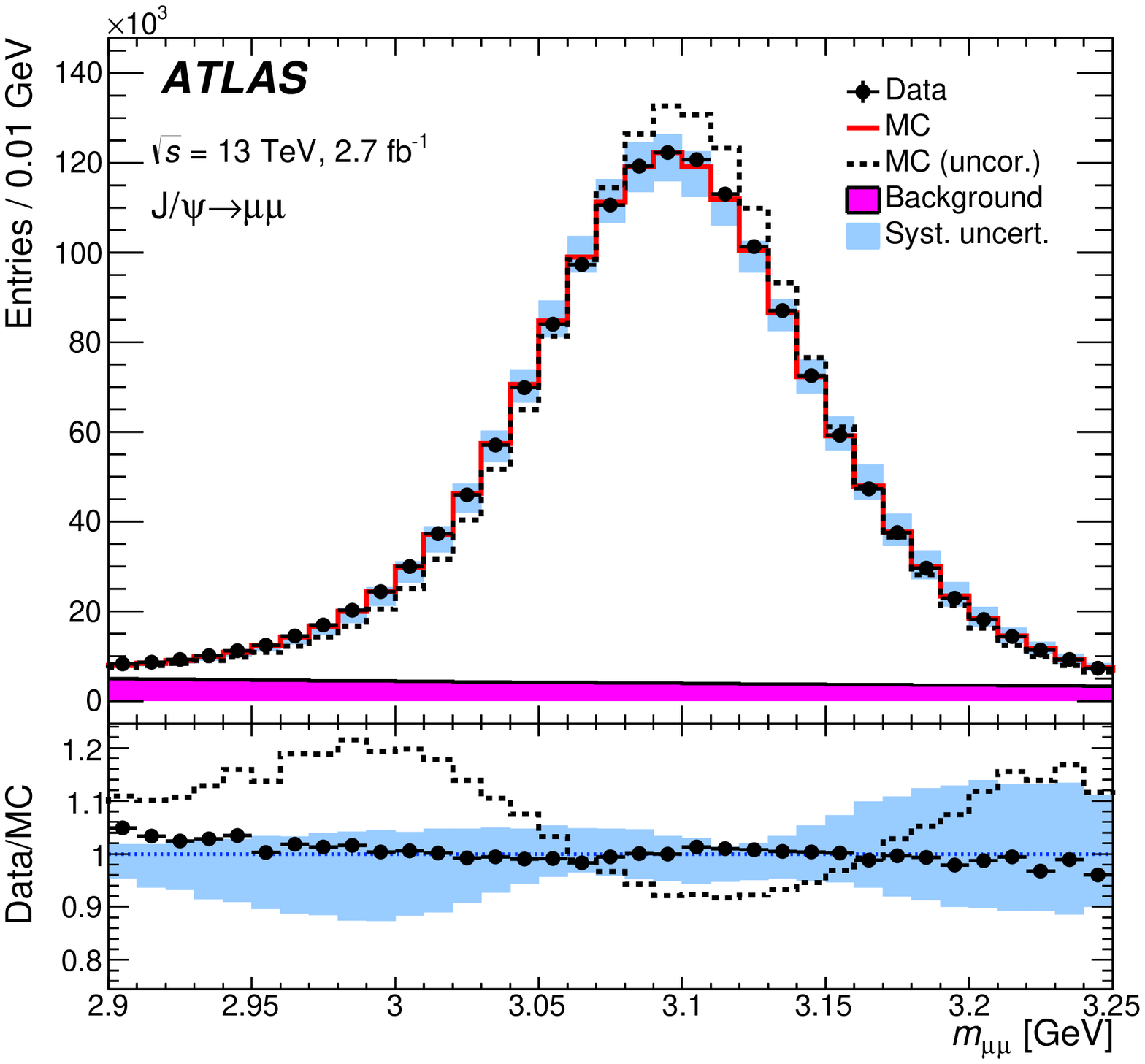}
    \caption{Dimuon invariant mass distribution of
  $Z\to\mu\mu$ (left) and  $J/\psi\to\mu\mu$  (right) candidate events reconstructed with CB muons.
 The upper panels show the invariant mass distribution for data and for the signal simulation plus the background estimate.
 The points show the data. The continuous line corresponds to the simulation with the MC momentum corrections applied
while the dashed lines show  the simulation when no correction is applied.
 Background estimates are added to the signal simulation. The band represents the effect of the systematic uncertainties on the MC momentum corrections.  The lower panels show the data to MC ratios.
 In the $Z$ sample, the MC background samples are added to the signal sample according to their expected cross sections.
 In the $J/\psi$ sample,  the background is estimated from a fit to the data as described in the text.
 The sum of background and signal MC distributions is normalised to the data. }
 \label{fig:Calib_InvMass}
 \end{figure*}

A better demonstration of the effectiveness of the momentum calibration is obtained by comparing, in data and simulation, the measurement of
the position  $m_{\mu\mu}$ and  resolution $\sigma_{\mu\mu}$ of the dimuon mass peaks,
extracted in bins of $\eta$ and $\pt$
from fits to the $\Jpsimm$  and  $\Zmm$ invariant mass distributions.

When the two muons have similar momentum resolution and angular effects are neglected,
the relative mass resolution, $\sigma_{\mu\mu}/m_{\mu\mu}$, is directly proportional to the relative muon momentum resolution,
$\sigma_{p_{\mu}}/p_{\mu}$:

\begin{equation}
  \frac{\sigma_{\mu\mu}}{m_{\mu\mu}} \, = \,  \frac{1}{\sqrt{2}} \frac{\sigma_{p_{\mu}}}{p_{\mu}}.
\label{eq:dmom}
\end{equation}
Similarly, the total muon momentum scale, defined as $s =  \langle (p^{\rm meas} - p^{\rm true})/  p^{\rm true} \rangle$,
is directly related to the dimuon mass scale, defined as  $ s_{\mu\mu} = \langle (m_{\mu\mu}^{\rm meas} - m_{\mu\mu}^{\rm true})/  m_{\mu\mu}^{\rm true} \rangle$:
\begin{equation}
 s_{\mu\mu} = \sqrt{s_{\mu_1}s_{\mu_2}},
\label{eq:defsmumu}
\end{equation}
 where $s_{\mu_1}$ and $s_{\mu_2}$ are the momentum scales of the two muons.

The dimuon mass resolution is obtained by fitting the  width of the invariant mass peaks.
In  $\Jpsimm$ decays, the intrinsic width of the resonance is negligible with respect to the experimental resolution.
The  bulk of the  peak is modelled by a Crystal Ball function; a  Gaussian distribution centred on the Crystal Ball function
is added to the signal description to model
the tails of the distribution.
The non-resonant background is described by an exponential function.
In $\Zmm$ decays, the fits use a convolution of the true lineshape (modelled by a Breit--Wigner function)
with an experimental resolution function (a combination of a Crystal Ball and a Gaussian function).
Similarly to the $J/\psi$, the non-resonant background is described by an exponential function.
The peak position and width of the Crystal Ball function are used as  estimators for the  $m_{\mu\mu}$ and $\sigma(m_{\mu\mu})$ variables
in the various $\eta$ and \pt\ bins.

Figure~\ref{fig:mass_eta} shows the  position of the peak of the invariant mass distribution, $ m_{\mu\mu}$, obtained from the fits to the $Z$ boson and  $J/\psi$  samples
as a function of the pseudorapidity of the highest-$\pt$ muon for CB pairs.
The distributions are shown for data as well as  corrected simulation, with the ratio of the two in the lower panel.
The simulation is in very good  agreement with the data.
Minor deviations are contained  within the scale systematic uncertainties of
$0.05\%$ in the barrel region, increasing with  $|\eta|$  to $0.1\% (0.3\%)$
 in the region $|\eta|\sim 2.5$ for $\Zmm$ ($\Jpsimm$) decays.
The systematic uncertainties shown in the plots include the effects of the uncertainties in the calibration constants described in Section~\ref{Sec:CorrectionExtraction} and the changes in the fit  parameterization.
The observed  level of agreement  demonstrates that the \pt\ calibration for combined muon tracks described above provides a very accurate
description of the momentum scale in all $\eta$ regions, over a wide  \pt\ range. Similar levels of data/MC agreement are observed for the ID and MS components of the combined tracks.

Figure~\ref{fig:width_eta} displays the dimuon mass resolution $\sigma(m_{\mu\mu})$ as a function of the leading-muon $\eta$ for the two resonances.
The dimuon mass resolution is about $1.2\%$ and $1.6\%$
at small $\eta$ values for $J/\psi$ and $Z$ bosons, respectively,
and increases to 1.6\% and 1.9\% in the endcaps.
This corresponds to a relative muon \pt\ resolution of  $1.7\%$ and $2.3\%$
 in the centre of the detector
and 2.3\% and 2.9\% in the endcaps for $J/\psi$ and $Z$ boson decays, respectively.
After applying the momentum corrections described above, the simulation reproduces the resolution measured in  data,
 well within the systematic uncertainties. The systematic uncertainties are estimated following the same procedure described
for the determination of the energy scale.
Good agreement between the dimuon mass resolution measured in data and simulation is also observed for the ID and MS components of the combined tracks.

The relative dimuon mass resolution  $\sigma_{\mu\mu}/m_{\mu\mu}$ depends approximately on  the average momentum of the muons, as shown in Eq.~(\ref{eq:dmom}).
This allows a direct comparison of the momentum resolution function determined with $J/\psi$ and $Z$ boson decays.
This is shown in Fig.~\ref{f:resol_vs_pt}, where the relative dimuon mass resolution from  $\Jpsimm$  and $\Zmm$ events is compared to simulation.
The $\Jpsimm$ and $\Zmm$ resolutions are in good agreement.
For the $J/\psi$, the average momentum is defined as $\langle \pt \rangle  = \frac{1}{2}(p_{\mathrm{T},1}+p_{\mathrm{T},2})$
 while for the $Z$ boson it is defined as
 \begin{equation}
 \pt^{*} = m_{Z} \sqrt{ \frac{\sin{\theta_1}\sin{\theta_2}}{2 (1-\cos \alpha_{12})}},
 \label{eq:angles}
 \end{equation}
\noindent where $m_{Z}$ is the $Z$ boson mass~\cite{pdg2014}, $\theta_1$ and $\theta_2$ are the polar angles of the two muons, and $\alpha_{12}$ is the opening angle of the muon pair.
 This definition, based on angular variables only, removes the correlation between the measurement of the dimuon mass and the average $\pt$.

\begin{figure*}[!hbt]
  \begin{center}
     \includegraphics[width=0.49\linewidth]{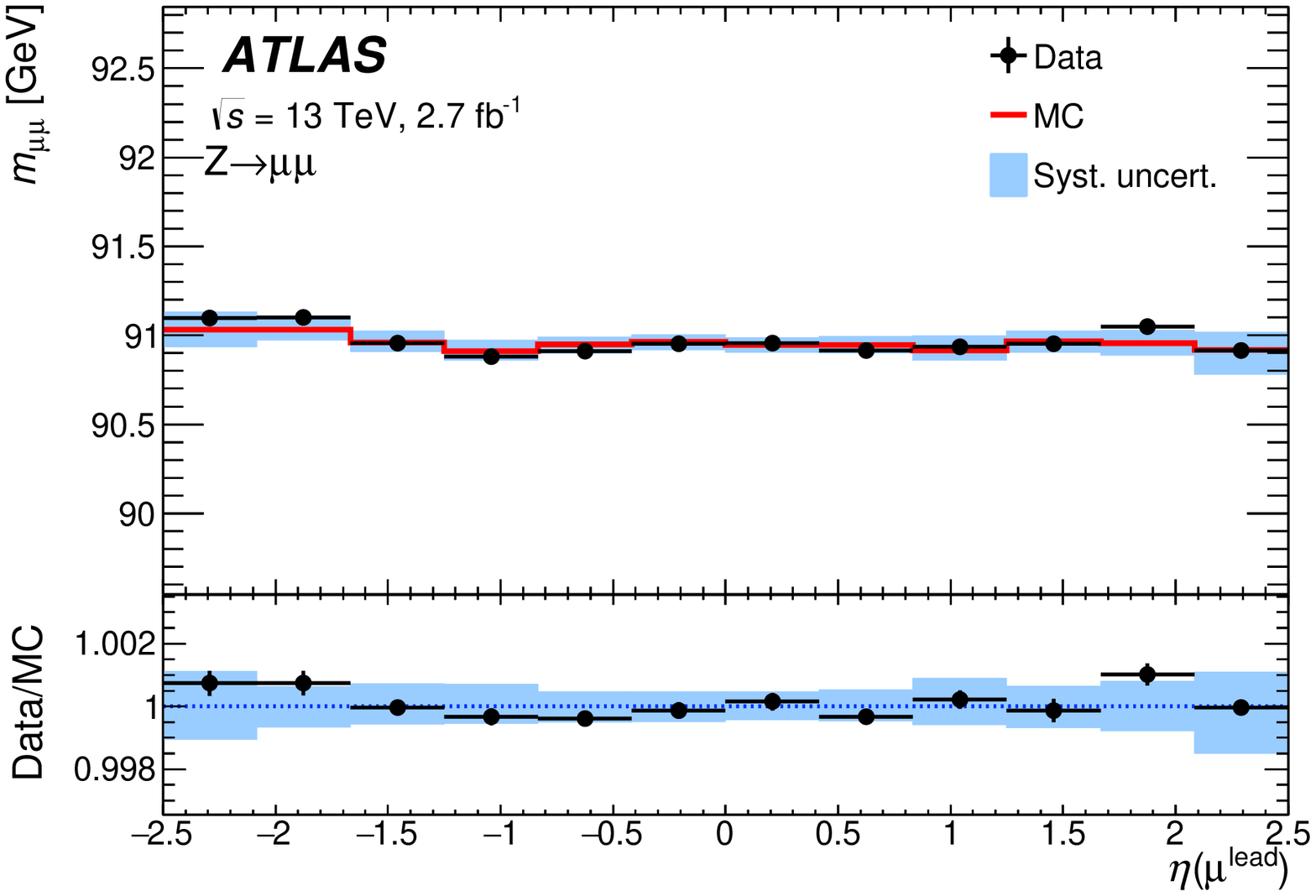}
     \includegraphics[width=0.49\linewidth]{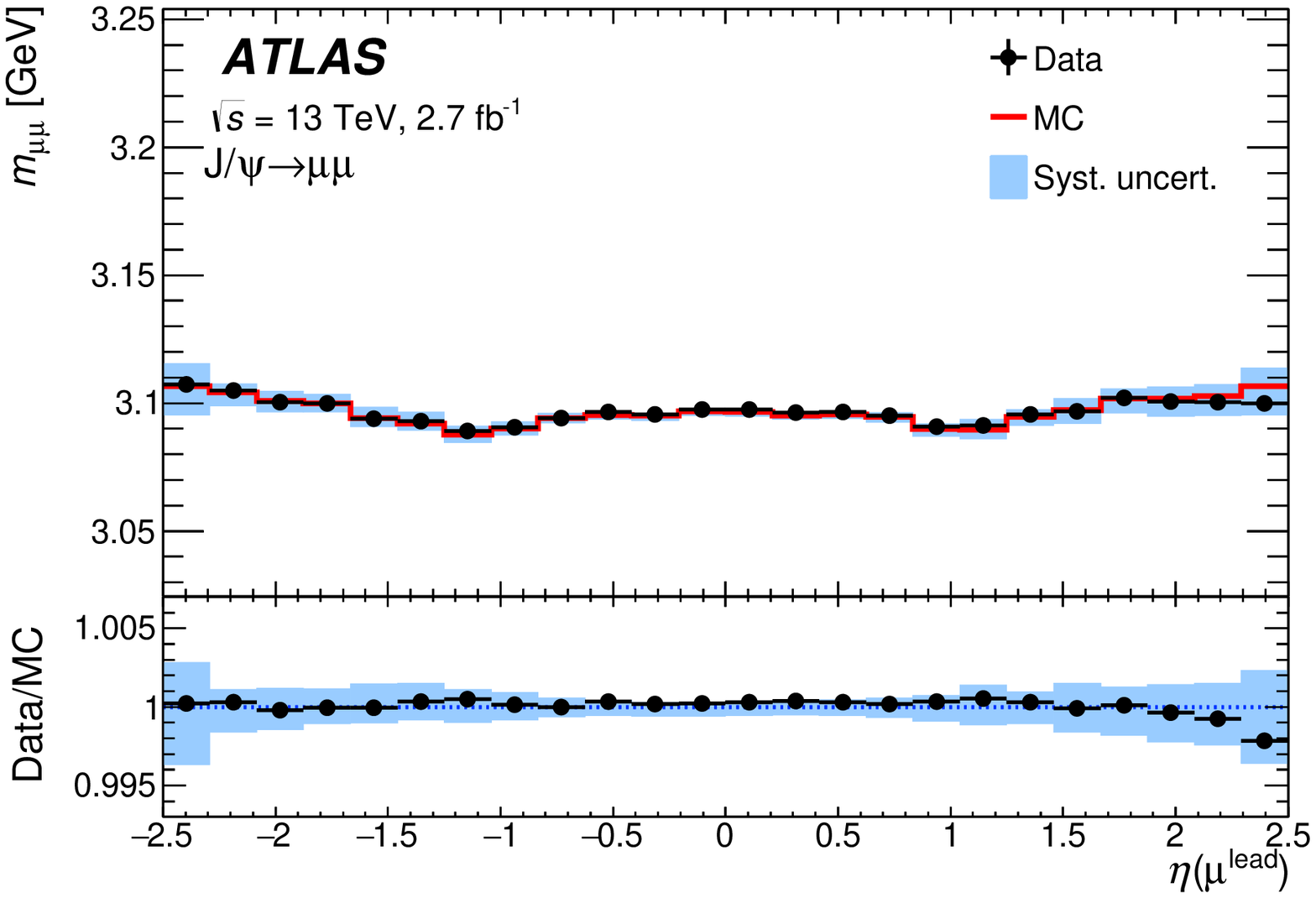}
    \caption{Fitted mean mass of the dimuon system for CB muons for $\Zmm$  (left) and  $\Jpsimm$ (right) events for
      data and corrected simulation as a function of the pseudorapidity of the highest-$\pt$ muon.
      The upper panels show the fitted mean mass value for data
      and corrected simulation.
The small variations of the invariant mass estimator as a function of pseudorapidity are
due to imperfect energy loss corrections and magnetic field description in the muon reconstruction.
Both effects are  well reproduced in the simulation.
      The lower panels show the data/MC ratio.
      The error bars represent the statistical uncertainty;  the shaded bands represent the systematic uncertainty in the correction and the systematic uncertainty in the extraction method added in quadrature.
}
\label{fig:mass_eta}
\end{center}
\end{figure*}

\begin{figure*}[!hbt]
  \begin{center}
     \includegraphics[width=0.49\linewidth]{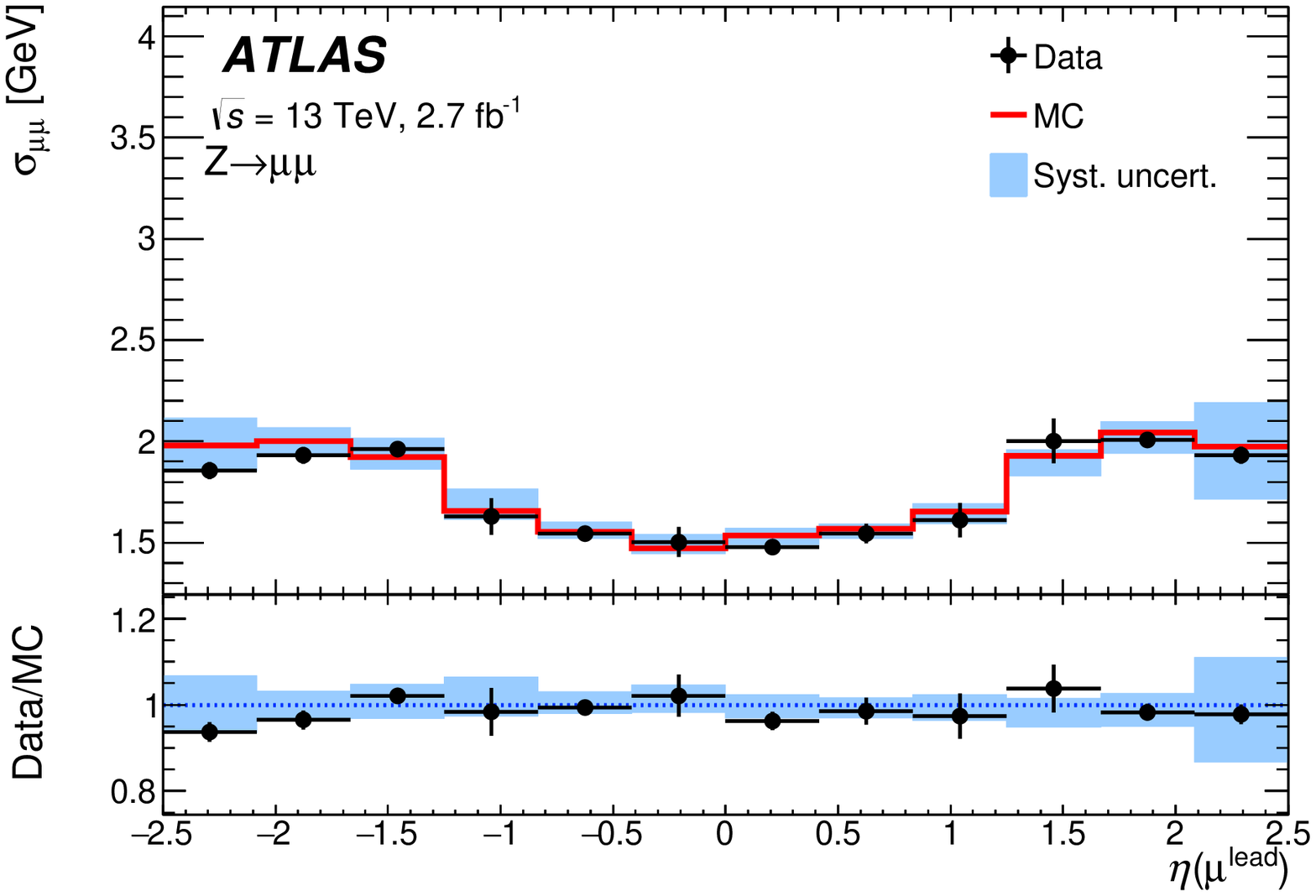}
     \includegraphics[width=0.49\linewidth]{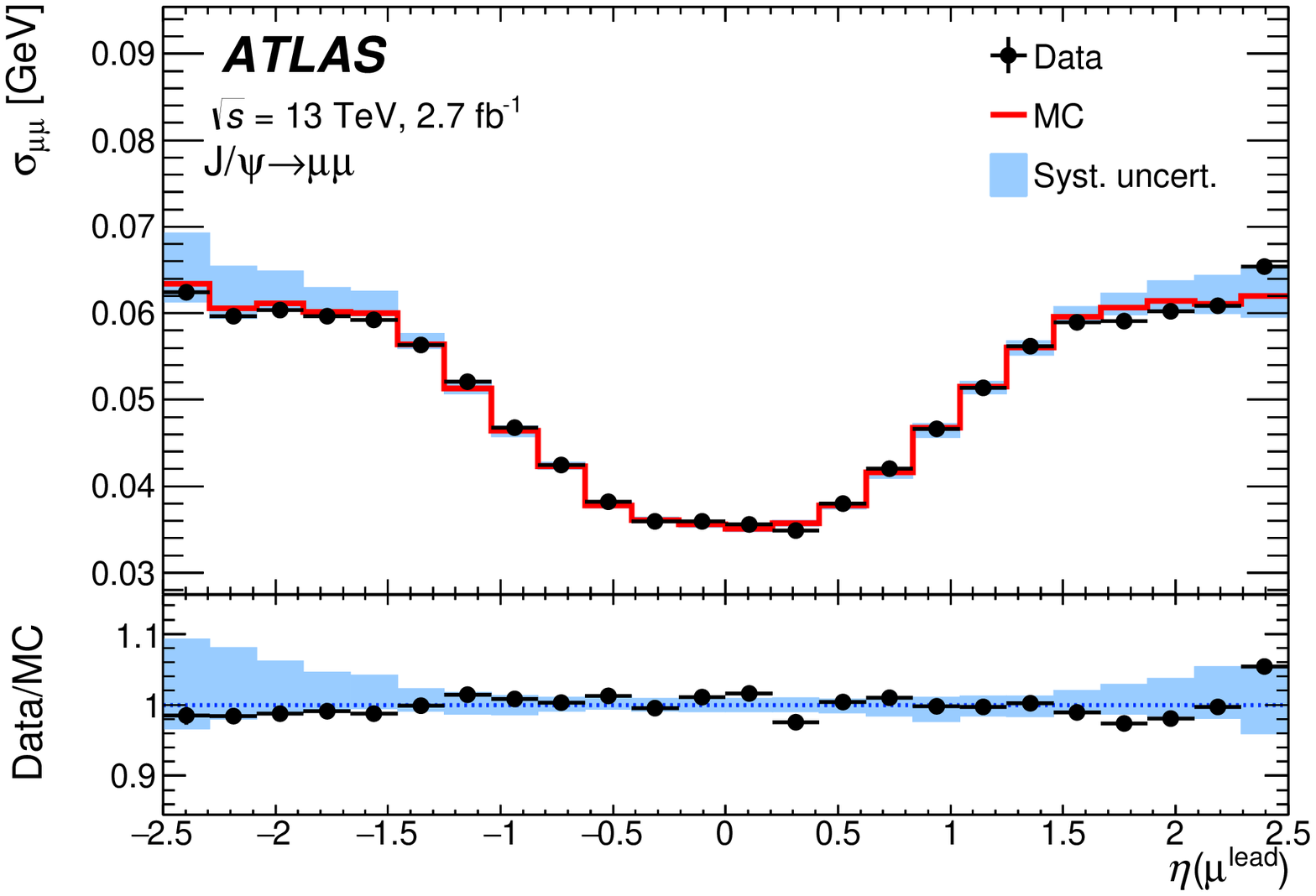}
    \caption{
    Dimuon invariant mass resolution for CB muons   for $\Zmm$  (left) and  $\Jpsimm$ (right) events for
      data and  corrected simulation as a function of the pseudorapidity of the highest-$\pt$ muon.
      The upper panels show the fitted resolution value for data
      and corrected simulation. The lower panels show the data/MC ratio.
      The error bars represent the statistical uncertainty;
      the shaded bands represent the systematic uncertainty in the correction and the systematic uncertainty
      in the extraction method added in quadrature.
}
\label{fig:width_eta}
\end{center}
\end{figure*}

\begin{figure*}[!htb]
\begin{center}
  \includegraphics[width=0.7\linewidth]{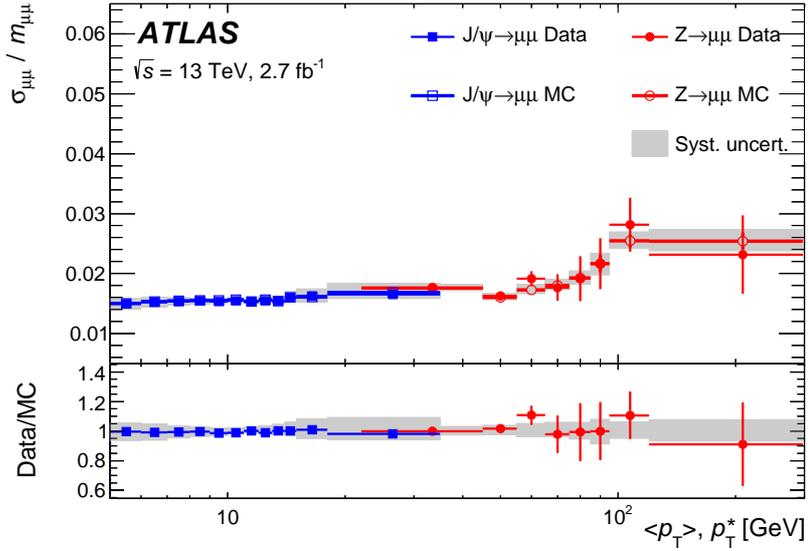}
 \end{center}
  \caption{
     Dimuon invariant mass resolution divided by the dimuon invariant mass for CB muons measured
    from  $\Jpsimm$ and  $\Zmm$ events as a  function of the
    average transverse momentum variables $\langle {p}_{\rm T} \rangle$ and  $ \pt^{*}$ defined in the text.
    Both muons are required to be in the same $|\eta|$ range.
    The error bars represent the statistical uncertainty while the bands show the systematic uncertainties.
}
   \label{f:resol_vs_pt}
\end{figure*}

\FloatBarrier

\section{Conclusions}
\label{sec:conclusions}

The performance of the ATLAS muon reconstruction has been measured using 3.2 fb$^{-1}$ of data
from  $pp$ collisions at $\sqrt{s} = 13$~\TeV\ recorded during the
 25 ns run at the LHC in 2015. A large calibration sample
consisting of
$\Zmm$ decays and
$\Jpsimm$ decays allows for a precise measurement  of the reconstruction and isolation efficiency as well as
of the momentum resolution and scale over a wide \pT\ range.

The muon reconstruction efficiency is close to $99\%$ over most of the pseudorapidity range of $|\eta|<2.5$ for $\pt>5$~\GeV.
The $\Zmm$  sample  enables a measurement of the efficiency
with a precision at the 0.2\% level for \pt\ $>$ 20~\GeV.
The $\Jpsimm$ sample  provides a measurement of the reconstruction efficiency between 5 and 20~\GeV\ with a
precision  better than 1\%.

The  $\Zmm$  sample is also used to measure the isolation efficiency for seven isolation  working points
in the  momentum range 10--120~\GeV.
The isolation efficiency varies between $93\%$ and $100\%$  depending on the selection and on the momentum of the particle,
and is well reproduced in the simulation.

The muon momentum scale and resolution have been studied in detail using  $\Jpsimm$ and  $\Zmm$ decays.
These studies are used to correct the simulation to improve the agreement with data
and to minimise the systematic uncertainties in physics analyses.
For $\Zmm$ decays, the uncertainty in the  momentum scale varies from a minimum of $0.05\%$ for \mbox{$|\eta| <1$} to a
maximum of \mbox{$0.3\%$} for  \mbox{$|\eta|\sim 2.5$}.
The dimuon mass resolution is about $1.2\%$ ($1.6\%$)
at small values of pseudorapidity for  $J/\psi$ ($Z$) decays, and increases
to 1.6\% and 1.9\% in the endcaps for $J/\psi$ and $Z$ decays, respectively.
This corresponds to a relative muon \pt\ resolution of  $1.7\%$ and $2.3\%$ at small values of pseudorapidity
and 2.3\% and 2.9\% in the endcaps for $J/\psi$ and $Z$ decays, respectively.
After applying momentum corrections, the  \pt\ resolution in data and simulation agree to better than 5\% for most of the \eta\ range.

\FloatBarrier
\section*{Acknowledgments}
\input{Acknowledgements}

\printbibliography


%

\newpage
\input{atlas_authlist}

\end{document}

%% file: Paper-metadata.tex
\AtlasTitle{Muon reconstruction performance of the ATLAS
  detector in  proton--proton collision data at $\sqrt{s}$=13~TeV}

\usepackage{authblk}
\makeatletter
\renewcommand\AB@affilsepx{, \protect\Affilfont}
\makeatother

\renewcommand\Authands{, } 
\renewcommand\Affilfont{\itshape\small} 
\AtlasRefCode{PERF-2015-10}

\PreprintIdNumber{CERN-EP-2016-033}

\AtlasJournalRef{\EPJC 76 (2016) 292}
\AtlasDOI{10.1140/epjc/s10052-016-4120-y}

\AtlasAbstract{\input{Abstract}}

\AtlasCoverSupportingNote{Support Note for 2015 Muon Combined Performance Paper}{http://cds.cern.ch/record/2052558/}
\AtlasCoverCommentsDeadline{18 Feb 2016}

\AtlasCoverAnalysisTeam{
C. Amelung,
G. Artoni,
G. Barone, 
K. Bachas,
M. Bellomo, 
F. Cardillo, 
M. Corradi, 
G. Cree, 
S. Dhaliwal, 
E. Diehl, 
C. Ferretti, 
M. Gignac, 
P-F. Giraud,
M. Goblirsch-Kolb,
H, Herde, 
S. Hu, 
P. Kluit,
N. Koehler, 
T. Koffas,
W. Leight, 
A. Lesage, 
K. Loew, 
J. Meyer, 
R. Nicolaidou,
I. Nomidis, 
N. Orlando,
G. Sciolla,
F. Sforza,
M. Sioli, 
N. Van Eldik,
S. Zambito, 
D. Zhang, 
L. Zhang
}

\AtlasCoverEdBoardMember{C. Guyot, S. Pagan-Griso, D. Orestano~(chair)}

\AtlasCoverEgroupEditors{atlas-perf-2015-10-editors@cern.ch}

\AtlasCoverEgroupEdBoard{atlas-perf-2015-10-editorial-board@cern.ch}



%% file: Abstract.tex
This article documents the performance of the ATLAS muon identification and reconstruction
using the  LHC dataset recorded at $\sqrt{s}$ = 13~\TeV\ in 2015. 
Using a large  sample of $J/\psi\to\mu\mu$ and $Z\to\mu\mu$  decays 
from 3.2 fb$^{-1}$ of $pp$ collision data, 
measurements of the reconstruction efficiency, as well as of the momentum
scale and resolution, are presented and compared to Monte Carlo simulations. 
The reconstruction efficiency is measured to be close to $99\%$  
over most of the covered phase space ($|\eta|<2.5$ and $5 <  p_{\mathrm{T}}   < 100$~\GeV). 
The isolation efficiency varies between $93\%$ and $100\%$  depending on the selection applied 
and on the momentum of the muon.
Both efficiencies are well reproduced in simulation. 
In the central region of the detector, the momentum resolution is measured to be $1.7\%$ ($2.3\%$) for muons from  $J/\psi\to\mu\mu$ ($Z\to\mu\mu$) decays, 
and the momentum scale is known with an uncertainty of $0.05\%$. 
In the region $|\eta |>2.2$, the $p_{\mathrm{T}} $ resolution for muons from $Z\to\mu\mu$  decays is  $2.9\%$ 
while  the precision of the momentum scale  for low-$p_{\mathrm{T}} $  muons from $J/\psi\to\mu\mu$ decays is about  $0.2\%$.

%% file: Acknowledgements.tex

We thank CERN for the very successful operation of the LHC, as well as the
support staff from our institutions without whom ATLAS could not be
operated efficiently.

We acknowledge the support of ANPCyT, Argentina; YerPhI, Armenia; ARC, Australia; BMWFW and FWF, Austria; ANAS, Azerbaijan; SSTC, Belarus; CNPq and FAPESP, Brazil; NSERC, NRC and CFI, Canada; CERN; CONICYT, Chile; CAS, MOST and NSFC, China; COLCIENCIAS, Colombia; MSMT CR, MPO CR and VSC CR, Czech Republic; DNRF and DNSRC, Denmark; IN2P3-CNRS, CEA-DSM/IRFU, France; GNSF, Georgia; BMBF, HGF, and MPG, Germany; GSRT, Greece; RGC, Hong Kong SAR, China; ISF, I-CORE and Benoziyo Center, Israel; INFN, Italy; MEXT and JSPS, Japan; CNRST, Morocco; FOM and NWO, Netherlands; RCN, Norway; MNiSW and NCN, Poland; FCT, Portugal; MNE/IFA, Romania; MES of Russia and NRC KI, Russian Federation; JINR; MESTD, Serbia; MSSR, Slovakia; ARRS and MIZ\v{S}, Slovenia; DST/NRF, South Africa; MINECO, Spain; SRC and Wallenberg Foundation, Sweden; SERI, SNSF and Cantons of Bern and Geneva, Switzerland; MOST, Taiwan; TAEK, Turkey; STFC, United Kingdom; DOE and NSF, United States of America. In addition, individual groups and members have received support from BCKDF, the Canada Council, CANARIE, CRC, Compute Canada, FQRNT, and the Ontario Innovation Trust, Canada; EPLANET, ERC, FP7, Horizon 2020 and Marie Sk{\l}odowska-Curie Actions, European Union; Investissements d'Avenir Labex and Idex, ANR, R{\'e}gion Auvergne and Fondation Partager le Savoir, France; DFG and AvH Foundation, Germany; Herakleitos, Thales and Aristeia programmes co-financed by EU-ESF and the Greek NSRF; BSF, GIF and Minerva, Israel; BRF, Norway; Generalitat de Catalunya, Generalitat Valenciana, Spain; the Royal Society and Leverhulme Trust, United Kingdom.

The crucial computing support from all WLCG partners is acknowledged
gratefully, in particular from CERN and the ATLAS Tier-1 facilities at
TRIUMF (Canada), NDGF (Denmark, Norway, Sweden), CC-IN2P3 (France),
KIT/GridKA (Germany), INFN-CNAF (Italy), NL-T1 (Netherlands), PIC (Spain),
ASGC (Taiwan), RAL (UK) and BNL (USA) and in the Tier-2 facilities
worldwide.

%% file: atlas_authlist.tex
\begin{flushleft}
{\Large The ATLAS Collaboration}

\bigskip

G.~Aad$^\textrm{\scriptsize 87}$,
B.~Abbott$^\textrm{\scriptsize 114}$,
J.~Abdallah$^\textrm{\scriptsize 65}$,
O.~Abdinov$^\textrm{\scriptsize 12}$,
B.~Abeloos$^\textrm{\scriptsize 118}$,
R.~Aben$^\textrm{\scriptsize 108}$,
M.~Abolins$^\textrm{\scriptsize 92}$,
O.S.~AbouZeid$^\textrm{\scriptsize 138}$,
N.L.~Abraham$^\textrm{\scriptsize 150}$,
H.~Abramowicz$^\textrm{\scriptsize 154}$
% ...
\bigskip
\\
% ...
$^{12}$ Institute of Physics, Azerbaijan Academy of Sciences, Baku, Azerbaijan\\
% ...
$^{65}$ University of Iowa, Iowa City IA, United States of America\\
% ...
$^{87}$ CPPM, Aix-Marseille Universit{\'e} and CNRS/IN2P3, Marseille, France\\
% ...
$^{92}$ Department of Physics and Astronomy, Michigan State University, East Lansing MI, United States of America\\
% ...
$^{108}$ Nikhef National Institute for Subatomic Physics and University of Amsterdam, Amsterdam, Netherlands\\
% ...
$^{114}$ Homer L. Dodge Department of Physics and Astronomy, University of Oklahoma, Norman OK, United States of America\\
% ...
$^{118}$ LAL, Univ. Paris-Sud, CNRS/IN2P3, Universit{\'e} Paris-Saclay, Orsay, France\\
% ...
$^{138}$ Santa Cruz Institute for Particle Physics, University of California Santa Cruz, Santa Cruz CA, United States of America\\
% ...
$^{150}$ Department of Physics and Astronomy, University of Sussex, Brighton, United Kingdom\\
% ...
$^{154}$ Raymond and Beverly Sackler School of Physics and Astronomy, Tel Aviv University, Tel Aviv, Israel\\
% ...
\end{flushleft}
